# A large calcium-imaging dataset reveals a systematic V4 organization for natural scenes

Tianye Wang[1-4,*], Haoxuan Yao[1-4,*], Tai Sing Lee[5,*], Jiayi Hong[1], Yang Li[1-4], Hongfei Jiang[1-4], Ian Max Andolina[6], Shiming Tang[1-4,✉]

[1]Peking University School of Life Sciences, Beijing 100871, China
[2]Peking-Tsinghua Center for Life Sciences, Beijing 100871, China
[3]IDG/McGovern Institute for Brain Research at Peking University, Beijing 100871, China.
[4]Key Laboratory of Machine Perception (Ministry of Education), Peking University, Beijing 100871, China
[5]Computer Science Department and Neuroscience Institute, Carnegie Mellon University, Pittsburgh, PA 15213, United States
[6]The Center for Excellence in Brain Science and Intelligence Technology, State Key Laboratory of Neuroscience, Key Laboratory of Primate Neurobiology, Institute of Neuroscience, Chinese Academy of Sciences, Shanghai 200031, China

[*]Equal contributions
Correspondence: tangshm@pku.edu.cn

## Summary

The visual system evolved to process natural scenes, yet most of our understanding of the topology and function of visual cortex derives from studies using artificial stimuli. To gain deeper insights into visual processing of natural scenes, we utilized widefield calcium-imaging of primate V4 in response to many natural images, generating a large dataset of columnar-scale responses. We used this dataset to build a digital twin of V4 via deep learning, generating a detailed topographical map of natural image preferences at each cortical position. The map revealed clustered functional domains for specific classes of natural image features. These ranged from surface-related attributes like color and texture to shape-related features such as edges, curvature, and facial features. We validated the model-predicted domains with additional widefield calcium-imaging and single-cell resolution two-photon imaging. Our study illuminates the detailed topological organization and neural codes in V4 that represent natural scenes.

## Introduction

Clustering neurons with similar properties into functional domains is an organizational principle of the brain. The investigation of this principle in the visual cortex has significantly contributed to our understanding of how the brain processes visual information. By identifying neuronal clusters with specific stimulus selectivities, we gain insights into the functional computations that are emphasized in a particular cortical area (Roe et al., 2012). For examples, the clustering of V1 orientation-selective neurons in orientation columns within a topological map might facilitate the computation of recurrent circuits for boundary completion (Bosking et al., 1997; Kapadia et al., 2000), whereas the neuronal clustering tuned to specific object-categories in IT can facilitate neural circuit computations for invariant object recognition (DiCarlo et al., 2012; Grill-Spector and Weiner, 2014). These functional organizational signatures can, therefore, highlight the computations carried out by neural circuits in each visual area.

To identify such functional clustering, researchers have used artificial stimuli that are parametric or have explicit semantic meaning to probe the visual cortex. While these experiments have uncovered numerous functional domains, our understanding of functional organization of the brain remains incomplete. The visual system has evolved to process natural scenes (Felsen and Dan, 2005), and artificial stimuli only cover a small part of the natural image space. Therefore, the functional organization revealed with artificial stimuli could miss important aspects of the functional organization of neural codes. These challenges are especially pronounced in intermediate visual area V4, because encoded features in V4 are not as simple as those found in the primary visual cortex, nor do they necessarily have explicit semantic meanings like those in IT cortex (Pasupathy et al., 2020).

To gain a more comprehensive understanding of the functional organization in visual cortex, it is important to study its structure within the context of natural images. However, studying functional organization based on natural stimuli has been challenging due to the complex structure and widespread dispersion of natural images within the high-dimensional image space (Rust and Movshon, 2005). Nevertheless, recent advances in neural recording techniques and deep-learning analysis methods offer promising solutions to overcome these hurdles. In this study, we achieved long-term stable widefield calcium-imaging (Seidemann et al., 2016) of neural responses at cortical columnar scale in awake monkeys. We captured the cortical responses spanning ten millimeters of the cortical surface of dorsal V4 in three monkeys, each to an extensive set of over 17,000 color natural images, at submillimeter resolution. The inclusion of a substantial and diverse collection of natural images provides more comprehensive information on V4 functional organization. This large-scale dataset allows us to train deep-learning models that can accurately predict the cortical response to arbitrary natural images (Allen et al., 2022; Bashivan et al., 2019; Ratan Murty et al., 2021; Richards et al., 2022). We then used these models to systematically investigate image feature preferences across the cortex. Through the combination of model predictions and additional experimental verification, we unveiled a finely organized V4 cortical preference map. The map comprises distinct functional domains, each preferring a specific variety of natural image features. These features range from surface-related attributes such as color and texture to shape-related features such as edge, curvature, and even components related to facial features.

## Results

**A large dataset of macaque V4 cortical responses to natural images**

We performed widefield calcium imaging to record V4 cortical responses to a large set of natural stimuli (Figure 1A). AAVs expressing the calcium indicator GCaMP5G (Chen et al., 2013) were injected into macaque visual cortical area V4. A 10mm-diameter optical window was implanted for imaging (Figure 1B). During calcium imaging, a 470 nm blue light was used to illuminate the cortex through the optical window, and the green fluorescence excitation signals were recording using a CCD camera (Camera A in Figure 1A).

To obtain neural responses to thousands of images we needed to integrate recordings across multiple days. However, we found that continuous blue light exposure causes severe photobleaching that results in a gradual attenuation of the fluorescence signal (Figure S1). To solve this problem, we developed an intermittent illumination paradigm. A shutter synchronized to the stimulus presentation was inserted into the blue light pathway to control the illumination. Each trial consisted of a 900 ms blank pre-stimulus period followed by 500 ms of stimulus presentation while the subject maintained fixation. Optical illumination lasting 250 ms occurred twice: one epoch 150 ms before, and one epoch 350 ms after stimulus onset. These corresponded to the baseline period before response initiation, and the peak response period, respectively (Figure 1C). The fluorescence images recorded in these two periods were used to calculate cortical responses (ΔF/F0, see Methods). This intermittent illumination method significantly improved the long-term stability of signals (Figure S1), facilitating the collection of the responses to thousands of images. To facilitate multi-day imaging registration, we recorded the blue reflectance images with another camera (Camera B in Figure 1A) and used the cortical capillaries (Figure 1B) as the reference for image registration. We found that the calcium signal was robust between different trials and across days (Figure 1D), allowing us to more confidently integrated data across multiple days into a large dataset.

We obtained the widefield calcium imaging dataset from the dorsal V4 of three monkeys. Each monkey's dataset includes a training set for fitting neural network models and a validation set for evaluating the model's prediction performance to novel stimuli. The stimulus presentation area was selected such that the receptive field of the imaged cortex, estimated using small grating parches, was positioned at the center of the stimulus (Figure S2B). The training set consists of single-trial cortical responses to 17,000-20,000 distinct color natural stimuli drawn from ImageNet (Deng et al., 2009) (see Figure 1E for examples). The validation set includes 500 additional natural images, 48 grating stimuli with varying orientations and spatial frequencies, and 8 distinct color patches (Figure S2C). Each validation stimulus was repeated ten times in a randomly interleaved fashion. Data collection for each monkey spanned six consecutive days. To monitor the stability of cortical responses across these days, in each recording day we tested 100 natural images selected from the validation set as fingerprint images. The high degree of correlation observed in the neural responses to fingerprint images across days provides further confirmation of the long-term stability of the measurements, supporting data integration (Figures S2D and S2E).

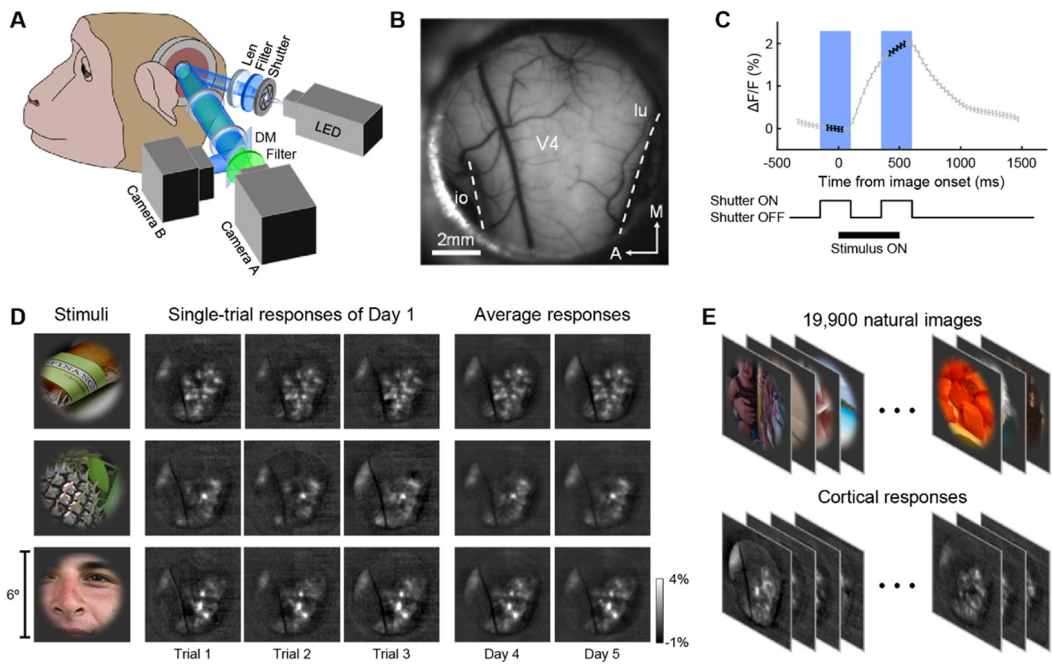

**Figure 1. Widefield calcium imaging of V4 cortical responses to natural scenes**
(A) Schematic of the widefield calcium imaging setup. A shutter was used to control intermittent illumination. On the detection side, a 525 nm dichroic mirror splits the reflectance light into green and blue, projecting them onto Camera A and Camera B, respectively.
(B) Example blue reflectance image recorded by Camera B in V4. A, anterior; M, medial; io, inferior occipital sulcus; lu, lunate sulcus.
(C) Top: Average time course of V4 responses to a stimulus, averaged over the responses to 100 natural images. Blue areas indicate "on" illumination. Image frames used for computing cortical responses are extracted from the periods indicated by the black labels on the response curve within the shutter "on" periods. Bottom: Control signal for shutter on-off; black bar denotes the stimulus presentation period.

(D) Example cortical responses to natural images across trials and days. The last two columns show the average responses of 5 repeats on Day 4 and Day 5.

(E) The dataset, used for training the deep neural network, contains cortical responses to 19,900 natural images from monkey C.

## DNN modelling of the cortical response dataset

We used the validation set to identify cortical regions that responded robustly to the visual stimuli (Figure S2A, See Methods). We focused our analysis on these regions and fitted deep learning models to capture the encoding relationship between stimuli and the cortical pixel's responses.

Earlier studies (Cadena et al., 2019; Schrimpf et al., 2020a; Yamins and DiCarlo, 2016; Yamins et al., 2014) suggest that DNNs optimized for object recognition provide a state-of-the-art model of the primate visual ventral stream. These task-driven DNN's internal representations can be used to fit neural responses to image stimuli via transfer learning. Typically, this is done by fitting the neural responses with a linear transform of feature activations of a specific DNN layer in response to the input image. This feature transfer approach has been shown to produce acceptable response prediction performance, particularly when the training dataset is of modest size (Allen et al., 2022; Cadena et al., 2019; Ratan Murty et al., 2021). One drawback of this approach however is that data fitting is restricted to the feature space of the pre-trained DNN, and can potentially fail to capture some of the characteristic of the feature space of the brain. One solution is to use the neural data to fine-tune the feature detectors upstream of the selected DNN layer via backpropagation. However, given the large number of trainable parameters in the DNN for recognition tasks, coupled with the modest size of the neural data, such models tend to overfit, which could account for the decrease in their generalization performance (Table S1). To address this issue, we employed a novel neural modelling strategy known as knowledge distillation (Gou et al., 2021; Wang and Yoon, 2021) to perform transfer learning (Figure 2A). Specifically, we used the feature transfer model mentioned above as the teacher to train a student network with significantly fewer parameters. Subsequently, the student network was fine-tuned on the targeted neural data again to obtain a final model that exhibit improved generalization performance.

We found this modelling strategy to be very effective in producing a model with superior neural response prediction performance on our measured data (Figure 2). We trained the feature transfer model with the Add-6 layer of ImageNet pre-trained Xception (Chollet, 2017) (Figures S3A and S3B). Knowledge distillation was then performed by training a small Xception-like DNN (Figure S3C, See Methods) on 100K image-response pairs predicted by the feature transfer model. The prediction performance of the transfer learning model obtained with knowledge distillation (KD Transfer, 73.1% of the achievable performance Figure 2B) is significantly better than the original feature transfer model (Feature Transfer, 68.2%; $p<10^{-9}$, Wilcoxon signed-rank test) and the small Xception-like DNN directly training with neural data (Direct, 68.8%; $p<10^{-9}$, Wilcoxon signed-rank test). The feature transfer model was better than the direct data-driven model with limited data, but this advantage dissipated when the data size reached 17K (Figure 2C). On the other hand, the KD Transfer model remains superior in prediction performance regardless of the size of the training dataset.

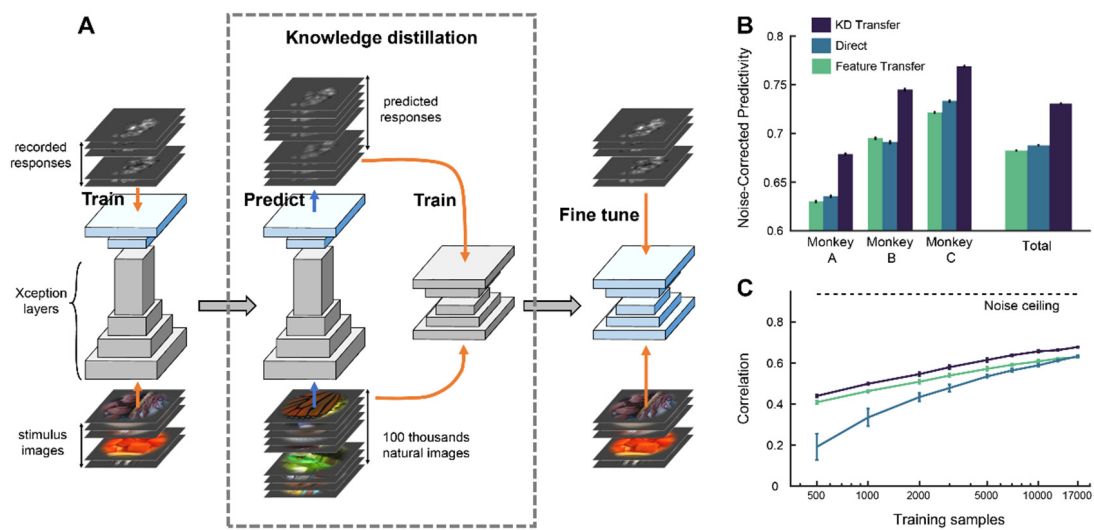

**Figure 2. DNN modeling on cortical response dataset**

(A) Schematic of transfer learning with knowledge distillation. We first used the neural data to train a feature transfer model which uses a two-layer perceptron to map the responses of the Add-6 layer of the Xception to input images to their evoked cortical responses. We then performed knowledge distillation to condense this feature-transfer model to a smaller DNN. This step was completed by training the small DNN using the responses of feature transfer model to 100K natural images. We finally fine-tuned the small DNN on the recorded neural dataset. Parameters of the network layers in blue are optimized on neural data.

(B) Neural response prediction performance of the feature transfer model (green), the data-driven model (Direct, light blue), and the knowledge distillation transfer model (KD Transfer, dark blue) on the data of 3 monkeys, averaged across all imaged cortical pixels. For each cortical pixel, the performance measure or predictivity of the model is quantified by computing the Pearson correlation between the predicted responses and recorded responses on validation images and then normalizing it with the noise ceiling of the pixel (see Methods, Figure S2F). The error bars denote the SEM. The KD Transfer model performed significantly better than the data-driven model and the feature transfer model ($p<10^{-9}$ in all cases; Wilcoxon signed-rank test), and the data-driven model performed significantly better than feature transfer model ($p<10^{-9}$; Wilcoxon signed-rank test) over the total of 7750 cortical pixels pooled across three monkeys.

(C) Performance measured in raw Pearson correlation as a function of training samples

relative to the noise ceiling is shown. Plotted lines and error bars indicate mean and standard deviation across results obtained from different bootstrap samples of the data.

**Natural image preference maps in V4**

Our neural network model that predicts V4 cortical responses with high accuracy essentially provides us with a digital twin of V4. This allows us to perform extensive tests *in silico* to dissect and characterize the neural coding in V4 (Abbasi-Asl et al., 2018; Bashivan et al., 2019; Ratan Murty et al., 2021; Ukita et al., 2019; Walker et al., 2019). A classic approach for this characterization identifies the optimal stimulus that elicits the strongest response in a neuron or pixel, or region (Bashivan et al., 2019; Ratan Murty et al., 2021; Walker et al., 2019). We therefore used our model to search for the optimal stimuli for each cortical pixel across a set of 50,000 natural images. We selected the top nine images preferred by the model for each location and visualized them in a 3 × 3 array over that location of the cortical surface (Figure 3A). The resulting map indicates the stimulus preference for each visually responsive pixel across the imaged area (Figure 3B, Supplementary Data). What appeared to be distinct clusters preferring different kinds of natural images were observed. To characterize the organizational structure of the map, we performed hierarchical clustering of cortical pixels based on the similarity of their top nine preferred images. The similarity in image preference between two cortical pixels was computed based on the Pearson correlation between the averaged cortical spatial response patterns to their respective preferred images (Figure S4, See Methods). Hierarchical clustering was then employed to group the cortical pixels into multiple domains, each showing a preference for natural images with similar characteristics (Figures 3C and 3D). Some domains emphasize specific colors, others prefer specific textures, some prefer specific transition boundaries, while others emphasize objects of specific shapes, such as round objects or even faces.

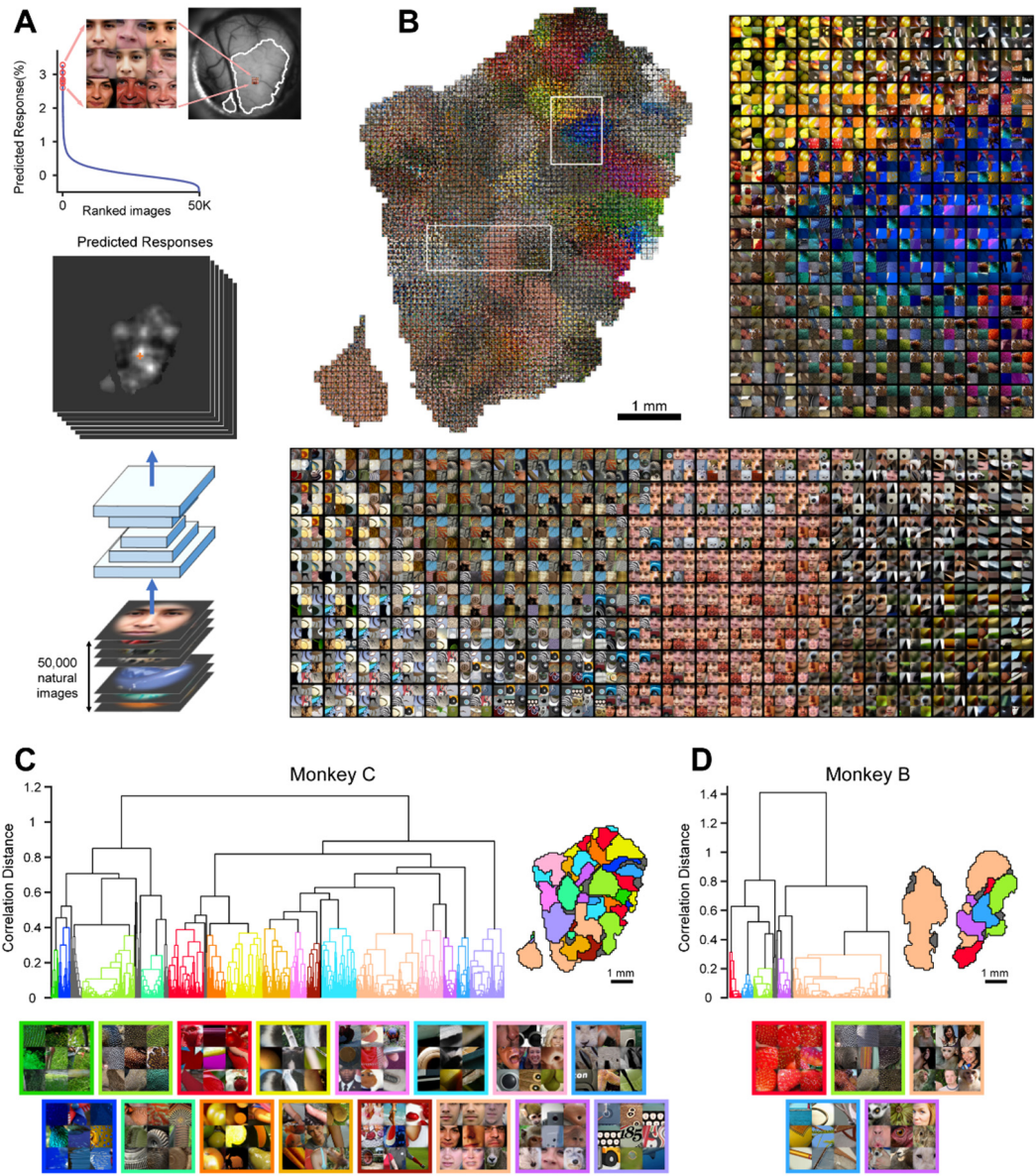

**Figure 3. Natural image preference maps predicted from a DNN model**

(A) The KD Transfer model was used to predict cortical responses to a 50,000-image set. For each cortical pixel (90×90μm physical size; the single pixel example is marked with an orange cross), the top 9 ranked images are shown as a 3×3 grid (only the center 4×4 degree of each image is visualized).

(B) Overall preference map obtained from monkey C. Domains preferring different colors can easily be observed. Zooming into the two regions marked by white rectangles, we observe cortical pixels preferring distinct shapes and surface attributes.

(C) The cortical pixels in monkey C were hierarchically clustered based on the similarity of the cortical responses to their top 9 preferred images. Clusters that contained connected region with more than 40 pixels were identified as functional domains, marked by distinct color. Clusters that did not meet the above criteria were marked in gray. Left: The dendrogram of the hierarchical clustering. Right: Cortical map with functional domains colored. Bottom: the images predicted to evoke strong responses for the identified domains. The same color scheme, indicating the functional domain categories, is used for the dendrogram, the cortical map and the image frames.

(D) Same as in (C), for monkey B.

To evaluate these model-predicted domains empirically, we performed additional widefield calcium imaging on monkey B and monkey C. For each domain, 16 preferred images predicted by the model were selected as test stimuli (Figures 4A and S5A). We found that stimuli selected for different domains elicited distinct cortical responses, and the measured activation patterns were consistent with model predictions (Figures 4A and 4B). Figure 4C shows the responses of each cortical pixel to each group of preferred images associated with the different functional domains (labeled in different color). We found the preferred image group based on their average response is the same as the image group associated the functional domain for 77.9% of the cortical pixels in monkey C and 87.3% of the cortical pixels for monkey B (Figure 4D). These results indicated that the functional domains revealed by the model-predicted preference map genuinely reflect the organization of image preferences in V4.

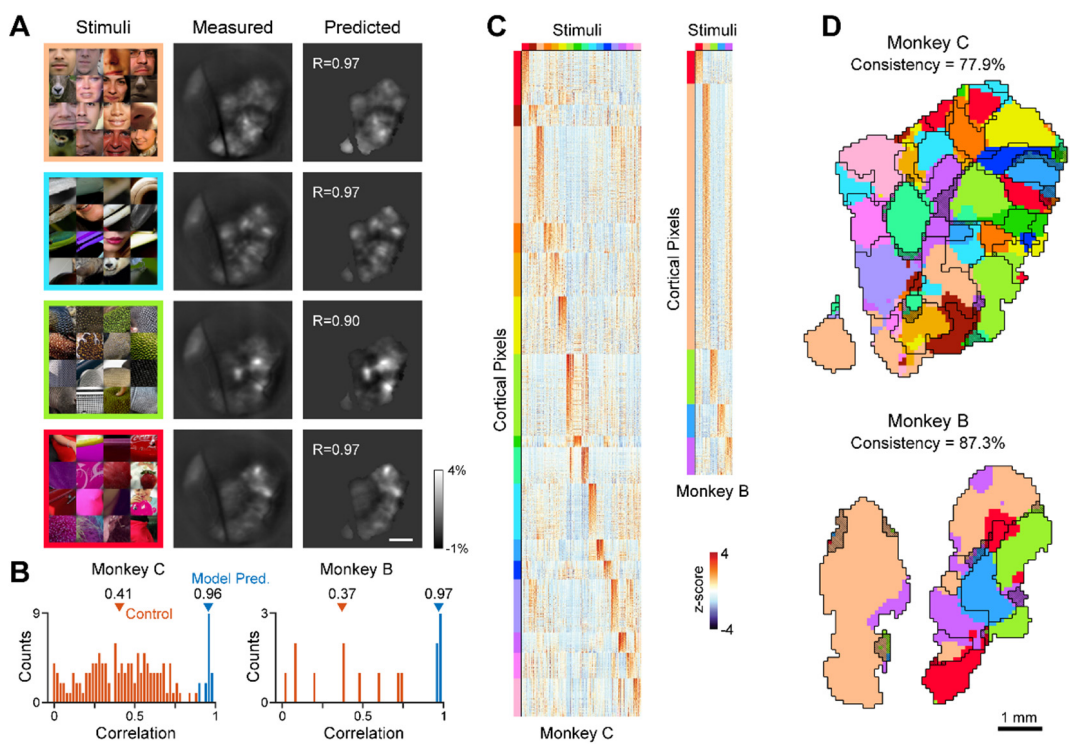

**Figure 4. Experimental verification of the model-predicted preference map**

(A) Test stimuli for the four example domains and their corresponding average activation patterns. The second and third columns show the measured and model-predicted average activation patterns, respectively. R denotes the correlation coefficient between the measured and the predicted pattern. The scale bar denotes 2mm.

(B) The blue bars in the histogram show the distribution of correlation coefficients between the measured and the model-predicted activation patterns. As a control, the red bars in the histogram display the correlation coefficients between measured activation patterns of all pairs of different stimulus categories. Arrows indicate means of the respective distributions.

(C) Population response matrices (z-scored, color scale lower left) to the test stimulus set for all classified cortical pixels (Figure 3C). Cortical pixels were sorted from top to bottom based on their responses to the test stimuli for their respective category.

(D) Measured stimulus preferences across the cortical surface. For each cortical pixel, we averaged its response to the test stimuli for each domain and identified the one with the highest response as its preferred category. The color of the cortical pixel represents its preferred category, the black contour outlines the model-predicted domains, and the hatched area represents the unclustered regions (grey in Figures 3C and 3D).

**Testing single-neuron selectivity on the preference map**

Having demonstrated a correspondence between the neural response at the columnar scale and the model-predicted preference map, we next tested the relationship between columnar-scale preference and single-neuron selectivity. We performed a series of two-photon calcium imaging recordings (Li et al., 2017) on ten selected fields of view (FOVs) of monkey C and monkey B respectively (Figures 5A, 5B and S6). To ensure that the test stimuli could effectively activated the neurons, we used the model to select a set of images preferred by dozens of representative cortical pixels to compose the test stimulus set (see Methods). The stimulus sets for monkey C and monkey B include 905 and 537 natural stimuli, respectively. We identified soma and dendrites that responded robustly to test stimuli as ROIs for the two-photon imaging analysis (see Methods, Figure S6B). We found that the cortical preferences obtained by widefield imaging are roughly consistent with the stimulus preferences of single neurons in the corresponding region (Figures 5C, 5D, and 5F). As shown in Figures 5C and 5D, the neurons at the face and dot domains also preferred face or dot stimuli. The average tuning of single-cell responses within each FOV is in good agreement with that measured by widefield imaging (Figures 5E and 5F). However, single cells (ROIs) exhibited a much greater degree of diversity and sparsity of tuning compared to the FOV responses, indicating that single neurons can encode and discriminate subtler variations of the image features (Figures 5C, 5D, and 5F).

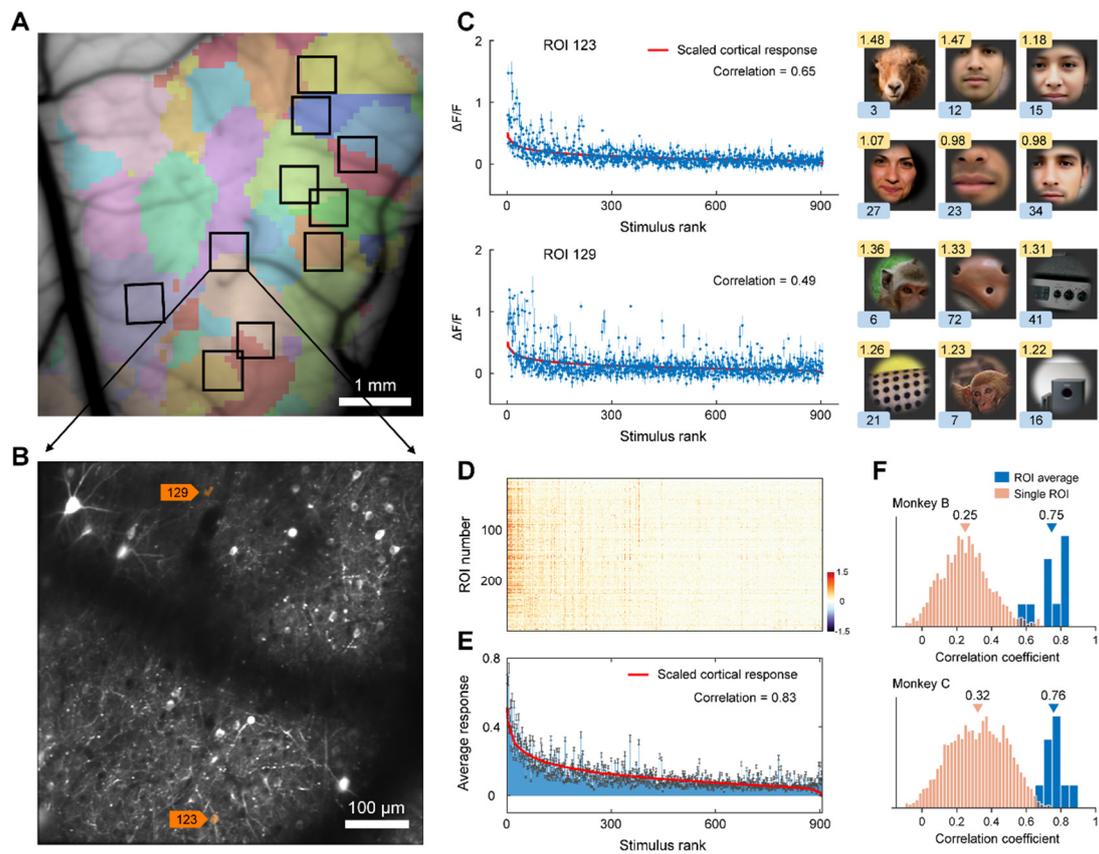

**Figure 5. Stimulus preference testing of single-neurons with two-photon imaging**

(A) Two-photon imaging recording sites in monkey C. The color hue represents the measured cortical preference, following the same color scheme as Figure 4D.

(B) The two-photon fluorescence image of an example field of view (FOV) averaged over all stimuli, located at the junction of the face and dot domains is shown.

(C) Responses of two example cell ROIs marked in (B) are shown. The error bar represents the SEM. The red line denotes the cortical responses of this FOV measured with widefield imaging. The stimuli are ranked according to the FOV's cortical responses. There is a significant correlation between the response of ROI and FOV, as shown. Stimuli with the top 6 neuronal responses are shown on the right side. For each stimulus, the number on the top left indicates its neuronal response, and the number on the bottom indicates its ranked stimulus index based on widefield imaging.

(D) Population response matrix of cell ROIs across the FOV region shown in (B).

(E) Average responses of all ROIs in the FOV region (error bar shows the SEM). As in (C), the red line denotes the FOV's response measured by widefield imaging, rescaled to minimize mean square error with the averaged ROI responses. The correlation between the two is 0.83.

(F) The blue histogram shows the distribution of correlation between the cortical response of each FOV measured by widefield imaging and the average responses of the ROIs within the FOV measured by two-photon imaging. The red histogram shows the distribution of correlation between the cortical response of each FOV and the single ROIs responses within the FOV. Arrows indicate means.

## Characterizing feature tuning using feature attribution analysis

Above, we characterized neural coding in terms of natural image preference. However, as natural images typically encompass a mixture of visual features, which specific features in the preferred images that are driving the neural responses remain unclear. To further characterize neural coding in terms of feature tuning, we performed a feature attribution analysis using SmoothGrad-Square (Hooker et al., 2019; Smilkov et al., 2017) method on our model. For a given input image, this gradient-based algorithm generates a heatmap that reflects the contribution of each pixel in the input image to the response of the target cortical region. The heatmap thus highlights the critical features in the image that drive the neural responses. Figure 6A shows several heatmap examples. For domains preferring dots, edges, and curvature, the critical features highlighted in resulting heatmaps align well with our presumed domain preference. Notably, for identified face domains, the heatmap reveals face components such as the nose and mouth are responsible for driving the neural responses.

We perform an *in vivo* wide-field imaging experiment to check whether the ablation of the critical features would indeed cause a significant drop in neural response. We targeted the face domain in monkey C and tested 12 sets of images, each derived from an image that was preferred by the face domain. Each set consisted of three test images: the original preferred image, the preferred image with the critical feature masked out, the preferred image with only the critical feature remaining (Figure 6B). The averaged response of the face domain to one set of test images is presented in Figure 6C. Figure 6D demonstrates the high consistency between the measured average cortical responses of the face domain and the model-predicted responses to the 12 sets of images (Pearson correlation = 0.84). We found that, although critical features constituted only a small part of the whole image, their occlusion resulted in a greater decrease in face domain's response compared to occlusion of all other parts of the image (Figure 6E). This evidence suggests that the critical feature revealed by the heatmap is indeed the part of the image critical for driving the response of the target domain.

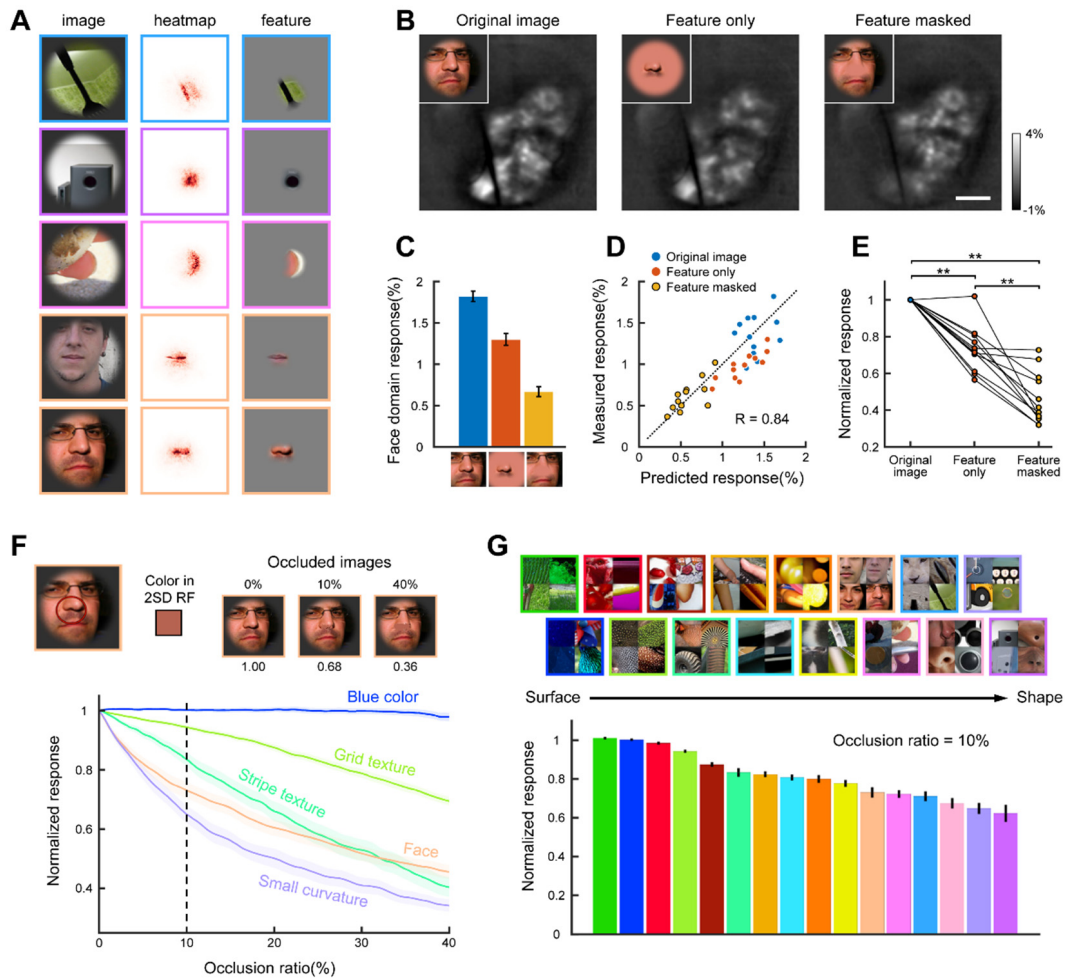

**Figure 6. Identifying critical image components with feature attribution analysis.**

(A) SmoothGrad-Square heatmaps of the preferred images highlight the critical features responsible for activating the domains. The first column shows the preferred image for example domains in monkey C; the second column shows the corresponding heatmap of each preferred image; the third column represents the image features emphasized by the heatmap. The box colors denote the domain categories.

(B) For the face domain in monkey C, we experimentally tested three types of stimuli – the original preferred images of the domain, images with only critical features visible, and images with critical features masked out. Cortical responses to a set of images are shown. The scale bar denotes 2 mm.

(C) Face domain's averaged responses to the example stimuli shown in **b**. The error bar represents the SEM.

(D) Face domain responses to the 12 sets of stimuli tested, showing a high correlation between the measured and the model predicted responses to the three types of stimuli across the different images.

(E) Comparison of face domain responses to the three types stimuli. P-values for paired t-tests are: original-image vs. feature-only, $1.35\times10^{-5}$; original-image vs. feature-masked, $5.85\times10^{-8}$; feature-only vs. feature-masked, $3.38\times10^{-4}$. P-values were corrected for multiple comparisons with Bonferroni correction.

(F) Model responses to example domains with their critical image regions occluded. For an image preferred by a domain, a fraction of image pixels estimated to be the most important based on the heatmap is occluded with the average color of the image within the 2 SD area of the receptive field (see Figure 7 for receptive field estimation). A face domain's preferred image with 0%, 20% and 40% occlusion and the normalized responses they evoked are shown below. Bottom: The relationship between the occlusion ratio and response for 5 example domains. The solid lines (±SEM) show the mean results of each domain's top 25 preferred images (in the 50,000-image set).

(G) Average responses of the top 25 images (normalized by the original responses, ±SEM) across different domains at 10% occlusion. Domains that prefer shape attributes exhibit a more significant drop in response under small occlusion compared to domains that prefer surface attributes.

We then performed more systematic testing of the stimuli with the critical features occluded for each identified functional domain on the model. For each test image, we used its average pixel color within the receptive field (RF) of the target domain to occlude the critical image region predicted by the heatmap (Figure 6F). This occlusion eliminates structured image patterns while preserving the average color information of the original image. We gradually increased the size of the occluded region, starting from the pixel with the highest heatmap value and replacing pixels successively in descending order of the heatmap value. We tested the responses of the domain to the preferred image as a function of the proportion of the pixels replaced relative to the domain's receptive field area (Figure 6F). We found that domains tuned to different features exhibited different degrees of sensitivity to the proportion of occlusion. Under a small proportion of occlusion (10% RF occluded), domains tuned to shape attributes, such as the face or curvature domains, exhibit a more significant response drop than domains tuned to texture and color. This might be attributed to the fact that shape information is represented by spatially precise pixels on the contour boundary, while surface features such as color and texture tend to involve large areas of uniform or homogeneous patterns. This difference in response to a small proportion of occlusion can be used as a metric to contrast the feature preference for surface and shape properties. The functional domains could be ordered according to this metric along a surface-shape axis (Figures 6G and S7A). Furthermore, functional domains preferring surface features tend to cluster together on the V4 map, and similarly for functional domains preferring shape features (Figure S7B). Such distinct clustering of functional domains might be a principle underlying the topological organization of V4.

The receptive field used in the above occlusion test was derived by averaging the heatmap of many natural images. Specifically, for a target cortical region, we averaged the heatmaps of its top 1,000 images from the 50,000-image set and fitted them with an elliptical Gaussian (Figures 7A and 7B). This method is applicable for estimating receptive fields of cortical regions with various feature preferences. We used this method to estimate the receptive field for each cortical pixel and obtained a fine-scale retinotopic map (Figures 7C and 7D). The average size of the receptive fields exhibited a linear relationship with eccentricity consistent with previous studies (Figure 7E) (Freeman and Simoncelli, 2011; Gattass et al., 1988).

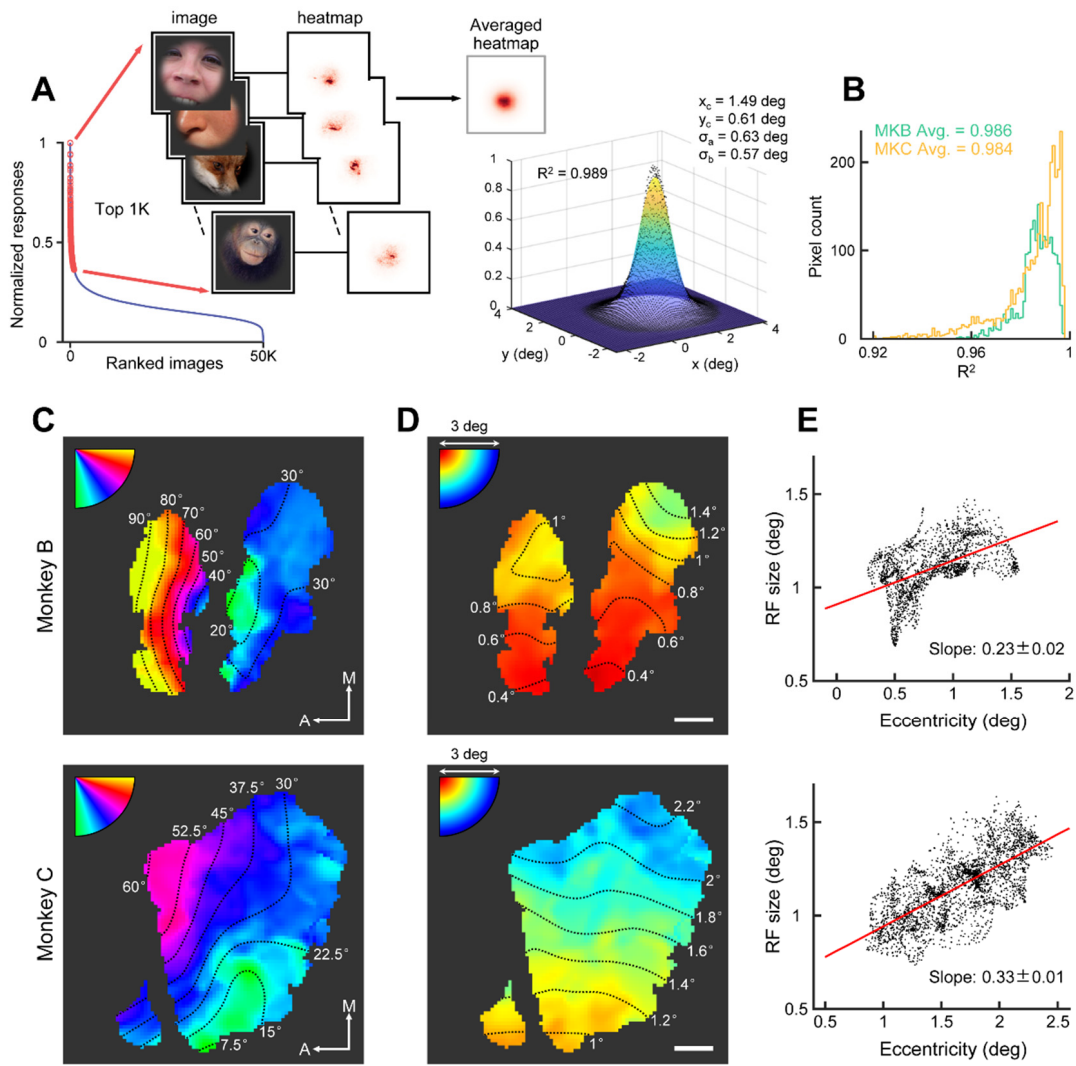

**Figure 7. Estimating receptive fields with averaged heatmaps.**
(A) Schematic of using heatmap to estimate receptive field (RF). For a target cortical site, we averaged the heatmaps of the top 1,000 images in the 50,000-image set and fitted them with an elliptical Gaussian.
(B) The distribution of R-square values obtained when fitting the averaged heatmap of cortical pixels with an elliptical Gaussian. The high R-square indicate that the elliptical Gaussian model provided a good fit to the averaged heatmap.
(C-D) The RF position maps of monkey B and monkey C in polar-angle(C) and eccentricity (D). These maps illustrate the spatial organization of receptive fields. A, anterior; M, medial. The scale bar denotes a distance of 1 mm.
(E) The relationship between eccentricity and RF size for cortical pixels in monkey B and monkey C. A linear regression was performed, revealing a significant linear relationship of RF size with eccentricity for monkey B ($F(1,1946) = 714$, $p = 2.99 \times 10^{-134}$) and monkey C ($F(1,3046) = 3171$, $p < 10^{-300}$).

# Discussion

Using widefield calcium imaging, we acquired a novel dataset of the cortical response of a large span of visual area V4, at 0.1 mm resolution, to tens of thousands of natural images. We used this dataset to train a deep learning model that can accurately predict the responses of every imaged cortical pixel. This V4 model allowed us to comprehensively characterize the neural tuning of cortical pixels in terms of natural stimuli. Using the model to identify preferred natural images for each cortical pixel, we identified a number of functional domains preferring distinct types of natural images. Further feature attribution analysis revealed that these domains could be arranged along a surface-shape axis based on their feature preferences, ranging from surface properties such as color and texture to shape attributes such as edge, curvature, and facial features.

**Using digital twins to uncover functional organization in V4**

Our deep learning model to accurately predict neural responses can be considered as a functional digital twin of V4. Using digital twins to characterize neural coding is an emerging research paradigm in visual neuroscience (Bashivan et al., 2019; Franke et al., 2022; Ratan Murty et al., 2021; Ustyuzhaninov et al., 2022; Walker et al., 2019). This paradigm liberates researchers from a number of normal constraints of neurophysiological experiments, allowing for systematic *in silico* exploration of neural tuning in the high-dimensional image space. Here we used the digital twin to search for images that elicited the strongest neural responses in a large set of natural images. This approach was not biased by any specific assumption about feature selectivity, mitigating the possibility of missing meaningful functional domains in the imaged region. Consequently, we were able to not only replicate previously reported functional domains that are tuned to color, orientation (Liu et al., 2020; Tanigawa et al., 2010), and curvature (Hu et al., 2020; Jiang et al., 2021; Tang et al., 2020) but also discover novel functional domains, including those that prefer facial features and texture. A study concurrent to ours (Willeke et al., 2023), adopted a similar deep-learning digital twin approach to study the functional organization in V4. In contrast to our investigation of the two-dimensional topological organization on the cortical surface using widefield calcium imaging, they focused on the columnar structure arranged orthogonally to the cortical surface using linear electrode arrays. They confirmed the presence of columnar structures in V4, as neurons arranged orthogonal to the cortical surface exhibited similar most preferred images. Remarkably, they also discovered clustering of neurons preferring textures and facial features, such as eyes, consistent with our findings. The convergence of our results and theirs provides additional confidence in the existence of these newly observed functional domains.

In addition to predicting *how* V4 would respond to arbitrary images, we also employed the digital twin to explore *what* attributes in a given image that are responsible for driving the neural responses by performing feature attribution analysis. Feature attribution analysis was developed by the artificial intelligence community to identify the features in the image responsible for a neural network's image classification decision. An occlusion-based feature attribution analysis, Randomized Input Sampling for Explanation (RISE) (Petsiuk et al., 2018), has been used to identify the critical features that drive the response of fMRI brain regions (Ratan Murty et al., 2021). We used an alternative gradient-based approach, called SmoothGrad-Square (Hooker et al., 2019), for heatmap calculation because we found that the mask used in the occlusion-based methods can itself be a feature. For example, zero-value mask is in fact an effective feature for domains preferring black bars. Hence, the occlusion-based methods can potentially give misleading results. Here, we showed that SmoothGrad-Square method provide a reasonable evaluation of critical features for the identified domains. By averaging the heatmaps produced by this method for many natural images, we obtain an estimate of the receptive fields of the neural units. This novel approach is particularly suitable for mapping the RF of neural units with complex feature selectivity, where traditional stimuli used for RF mapping may be unable to elicit a neural response. Feature attribution analysis also revealed that the feature preferences of the identified domains might lie along a surface-shape axis. A number of earlier studies had reported that V4 neurons were selective for shape (Carlson et al., 2011; Gallant et al., 1993; Nandy et al., 2013; Pasupathy and Connor, 2001; Pasupathy and Connor, 2002) or surface properties (Arcizet et al., 2008; Bushnell et al., 2011; Kim et al., 2022; Okazawa et al., 2014; Zeki, 1973). A study on the joint coding of shape and texture (Kim et al., 2019) found that V4 neurons lie along a continuum from strong tuning for boundary curvature of shapes to strong tuning for perceptual texture dimensions. Our results support these ideas and identify neurons tuned to shape properties or surface properties are organized topologically in distinct clusters in V4. The strong responses of neurons tuned to surface properties are always driven by spatially dispersed features, while the strong responses of neurons tuned to shape properties are always driven by spatially concentrated features. There is a potential need for separate and specialized neural circuits for computing these distinct feature tunings. Separating the surface and shape processing units into two distinct clusters might facilitate these computations.

**Knowledge distillation as a strategy for creating high-performance digital twins**

A model that can accurately predict neural responses is crucial for this study. To fully utilize the large dataset to improve model performance, we adopted a novel neural modeling strategy based on knowledge distillation. Knowledge distillation is a transfer learning method that compresses a large model and transfers its knowledge to a small one (Gou et al., 2021; Wang and Yoon, 2021). This method enables the fine-tuning of the features of a pre-trained network using neural data, resulting in better performance than earlier methods that used a linear combination of pre-trained features for transfer learning. It would be interesting to further investigate how this knowledge distillation strategy could work on modeling different modality neural data from V1, V2, IT.

**The utility of our widefield calcium imaging dataset**

Visual neuroscience is currently facing a pressing need for large datasets that sample neural responses to diverse natural scenes (Naselaris et al., 2021). Our study not only adds to the scientific knowledge in this area but also presents a novel and substantial V4 dataset acquired through widefield calcium imaging. This dataset holds the potential to serve as a valuable benchmark for testing and evaluating computational models of the brain. A popular strategy of brain modeling is to develop task-driven artificial neural networks so that their network features can best explain the relationship between stimuli and neural responses (Chang et al., 2021; Schrimpf et al., 2020b; Zhuang et al., 2021). The quality and richness of the neural data available in our dataset will allow a reliable evaluation of the competing computational models in terms of neural response prediction and representational similarity. Furthermore, our calcium imaging dataset provides information about columnar-scale topographic organization, which is not available in existing datasets based on fMRI (Allen et al., 2022) or electrophysiological recordings (Cadena et al., 2019; Majaj et al., 2015). This unique feature of our dataset makes it particularly valuable as a benchmark for assessing computational models that explore the principles underlying functional organization of the primate visual cortex (Blauch et al., 2022; Durbin and Mitchison, 1990; Koulakov and Chklovskii, 2001).

**Limitations and future directions**

Our study uncovered V4 functional domains representing rich natural image features, and the feature preferences of these domains can be arranged along a surface-shape axis. These findings were predominantly based on data from two monkeys (monkey B and monkey C), who have receptive fields near the fovea. The receptive fields of a third monkey were more peripheral, and our tested stimuli were not large enough to cover the entire receptive fields of the imaged area in this monkey. Hence, we could not be certain of capturing the complete properties of their neural tunings. Further study is therefore required to elucidate the potential difference in neural codes between the foveal and the peripheral visual fields.

We clustered cortical pixels based on the similarity of their most-preferred images. Although images evoking the strongest cortical response represent an important aspect of neural selectivity, they only provide partial characterization of the functions of individual neurons or neuronal populations. Neurons may share a most preferred image but exhibit different tuning responses to other images. To gain a more comprehensive understanding of functional organization, it is necessary to characterize cortical function in terms of the complete tuning responses. This would involve analyzing how neurons respond to a wide range of natural images and how they encode the statistical regularities in natural scenes. An accurate digital twin can be useful for such future study.

Our two-photon imaging confirm that the widefield imaging signals correspond to the average neuronal population responses at the submillimeter scale. However, the two-photon imaging also reveals that individual neuron's selectivity is more complex and diverse than the widefield imaging signals, indicating that single neurons carry more detailed coding information. A coarse-to-fine approach that uses widefield calcium imaging to comprehensively characterize the functional map of a cortical region, followed by two-photon imaging targeting regions of interest to study the tuning properties of neurons, will be an effective strategy for deciphering the function of neurons within their response populations.

## Acknowledgments

We would like to thank Doris Tsao and Winrich A. Freiwald for their helpful comments and suggestion on the manuscript. We thank the Peking University Laboratory Animal Center for excellent animal care. This work was supported by National Natural Science Foundation of China (grant no. 31730109 and U1909205) and funds from the Peking-Tsinghua Center for Life Sciences. This work was also supported by High-performance Computing Platform of Peking University.

## Author contributions

Conceptualization, T.W., T.S.L, and S.T.; Methodology, T.W., H.Y., and S.T.; Software, T.W., H.Y., J.H., and S.T.; Formal Analysis, T.W., H.Y., J.H., and S.T.; Investigation, H.Y. and T.W.; Resources, Y.L., H.J., and S.T.; Writing-Original Draft, T.W., T.S.L., H.Y., and S.T.; Writing-Review & Editing, T.S.L., I.M.A, T.W., and S.T.; Visualization, T.W., H.Y., J.H., and T.S.L.; Supervision, S.T.; Funding Acquisition, S.T.

## Declaration of interests

The authors declare no competing interests.

## References

Abbasi-Asl, R., Chen, Y., Bloniarz, A., Oliver, M., Willmore, B.D., Gallant, J.L., and Yu, B. (2018). The DeepTune framework for modeling and characterizing neurons in visual cortex area V4. bioRxiv,

465534.

Allen, E.J., St-Yves, G., Wu, Y., Breedlove, J.L., Prince, J.S., Dowdle, L.T., Nau, M., Caron, B., Pestilli, F., Charest, I., *et al.* (2022). A massive 7T fMRI dataset to bridge cognitive neuroscience and artificial intelligence. Nat Neurosci *25*, 116-126.

Arcizet, F., Jouffrais, C., and Girard, P. (2008). Natural textures classification in area V4 of the macaque monkey. Experimental Brain Research *189*, 109-120.

Bashivan, P., Kar, K., and DiCarlo, J.J. (2019). Neural population control via deep image synthesis. Science *364*.

Blauch, N.M., Behrmann, M., and Plaut, D.C. (2022). A connectivity-constrained computational account of topographic organization in primate high-level visual cortex. Proc Natl Acad Sci U S A *119*.

Bosking, W.H., Zhang, Y., Schofield, B., and Fitzpatrick, D. (1997). Orientation selectivity and the arrangement of horizontal connections in tree shrew striate cortex. Journal of Neuroscience *17*, 2112-2127.

Bushnell, B.N., Harding, P.J., Kosai, Y., Bair, W., and Pasupathy, A. (2011). Equiluminance Cells in Visual Cortical Area V4. Journal of Neuroscience *31*, 12398-12412.

Cadena, S.A., Denfield, G.H., Walker, E.Y., Gatys, L.A., Tolias, A.S., Bethge, M., and Ecker, A.S. (2019). Deep convolutional models improve predictions of macaque V1 responses to natural images. PLoS Comput Biol *15*, e1006897.

Carlson, E.T., Rasquinha, R.J., Zhang, K., and Connor, C.E. (2011). A Sparse Object Coding Scheme in Area V4. Current Biology *21*, 288-293.

Chang, L., Egger, B., Vetter, T., and Tsao, D.Y. (2021). Explaining face representation in the primate brain using different computational models. Curr Biol *31*, 2785-2795 e2784.

Chen, T.W., Wardill, T.J., Sun, Y., Pulver, S.R., Renninger, S.L., Baohan, A., Schreiter, E.R., Kerr, R.A., Orger, M.B., Jayaraman, V., *et al.* (2013). Ultrasensitive fluorescent proteins for imaging neuronal activity. Nature *499*, 295-300.

Chollet, F. (2017). Xception: Deep Learning with Depthwise Separable Convolutions. 30th Ieee Conference on Computer Vision and Pattern Recognition (Cvpr 2017), 1800-1807.

Deng, J., Dong, W., Socher, R., Li, L.J., Li, K., and Li, F.F. (2009). ImageNet: A Large-Scale Hierarchical Image Database. Proc Cvpr Ieee, 248-255.

DiCarlo, J.J., Zoccolan, D., and Rust, N.C. (2012). How does the brain solve visual object recognition? Neuron *73*, 415-434.

Durbin, R., and Mitchison, G. (1990). A dimension reduction framework for understanding cortical maps. Nature *343*, 644-647.

Felsen, G., and Dan, Y. (2005). A natural approach to studying vision. Nature Neuroscience *8*, 1643-1646.

Franke, K., Willeke, K.F., Ponder, K., Galdamez, M., Zhou, N., Muhammad, T., Patel, S., Froudarakis, E., Reimer, J., Sinz, F.H., *et al.* (2022). State-dependent pupil dilation rapidly shifts visual feature selectivity. Nature *610*, 128-134.

Freeman, J., and Simoncelli, E.P. (2011). Metamers of the ventral stream. Nature Neuroscience *14*, 1195-1201.

Gallant, J.L., Braun, J., and Vanessen, D.C. (1993). Selectivity for Polar, Hyperbolic, and Cartesian Gratings in Macaque Visual-Cortex. Science *259*, 100-103.

Gattass, R., Sousa, A.P.B., and Gross, C.G. (1988). Visuotopic Organization and Extent of V3 and V4

of the Macaque. Journal of Neuroscience *8*, 1831-1845.

Giovannucci, A., Friedrich, J., Gunn, P., Kalfon, J., Brown, B.L., Koay, S.A., Taxidis, J., Najafi, F., Gauthier, J.L., Zhou, P., et al. (2019). CalmAn an open source tool for scalable calcium imaging data analysis. eLife *8*.

Gou, J., Yu, B., Maybank, S.J., and Tao, D. (2021). Knowledge distillation: A survey. International Journal of Computer Vision *129*, 1789-1819.

Grill-Spector, K., and Weiner, K.S. (2014). The functional architecture of the ventral temporal cortex and its role in categorization. Nat Rev Neurosci *15*, 536-548.

Hooker, S., Erhan, D., Kindermans, P.-J., and Kim, B. (2019). A benchmark for interpretability methods in deep neural networks. Advances in neural information processing systems *32*.

Hu, J.M., Song, X.M., Wang, Q., and Roe, A.W. (2020). Curvature domains in V4 of macaque monkey. Elife *9*.

Jiang, R., Andolina, I.M., Li, M., and Tang, S. (2021). Clustered functional domains for curves and corners in cortical area V4. Elife *10*.

Kapadia, M.K., Westheimer, G., and Gilbert, C.D. (2000). Spatial distribution of contextual interactions in primary visual cortex and in visual perception. J Neurophysiol *84*, 2048-2062.

Kim, T., Bair, W., and Pasupathy, A. (2019). Neural Coding for Shape and Texture in Macaque Area V4. The Journal of Neuroscience *39*, 4760-4774.

Kim, T., Bair, W., and Pasupathy, A. (2022). Perceptual Texture Dimensions Modulate Neuronal Response Dynamics in Visual Cortical Area V4. The Journal of Neuroscience *42*, 631-642.

Kingma, D.P., and Ba, J. (2014). Adam: A method for stochastic optimization. arXiv preprint arXiv:14126980.

Koulakov, A.A., and Chklovskii, D.B. (2001). Orientation preference patterns in mammalian visual cortex: a wire length minimization approach. Neuron *29*, 519-527.

Li, M., Liu, F., Jiang, H., Lee, T.S., and Tang, S. (2017). Long-Term Two-Photon Imaging in Awake Macaque Monkey. Neuron *93*, 1049-1057 e1043.

Liu, Y., Li, M., Zhang, X., Lu, Y., Gong, H., Yin, J., Chen, Z., Qian, L., Yang, Y., Andolina, I.M., et al. (2020). Hierarchical Representation for Chromatic Processing across Macaque V1, V2, and V4. Neuron *108*, 538-550 e535.

Majaj, N.J., Hong, H., Solomon, E.A., and DiCarlo, J.J. (2015). Simple Learned Weighted Sums of Inferior Temporal Neuronal Firing Rates Accurately Predict Human Core Object Recognition Performance. J Neurosci *35*, 13402-13418.

Nandy, Anirvan S., Sharpee, Tatyana O., Reynolds, John H., and Mitchell, Jude F. (2013). The Fine Structure of Shape Tuning in Area V4. Neuron *78*, 1102-1115.

Naselaris, T., Allen, E., and Kay, K. (2021). Extensive sampling for complete models of individual brains. Current Opinion in Behavioral Sciences *40*, 45-51.

Okazawa, G., Tajima, S., and Komatsu, H. (2014). Image statistics underlying natural texture selectivity of neurons in macaque V4. Proceedings of the National Academy of Sciences *112*.

Pasupathy, A., and Connor, C.E. (2001). Shape representation in area V4: Position-specific tuning for boundary conformation. J Neurophysiol *86*, 2505-2519.

Pasupathy, A., and Connor, C.E. (2002). Population coding of shape in area V4. Nature Neuroscience *5*, 1332-1338.

Pasupathy, A., Popovkina, D.V., and Kim, T. (2020). Visual Functions of Primate Area V4. Annual Review of Vision Science *6*, 363-385.

Petsiuk, V., Das, A., and Saenko, K. (2018). RISE: Randomized Input Sampling for Explanation of Black-box Models.

Ratan Murty, N.A., Bashivan, P., Abate, A., DiCarlo, J.J., and Kanwisher, N. (2021). Computational models of category-selective brain regions enable high-throughput tests of selectivity. Nat Commun *12*, 5540.

Richards, B., Tsao, D., and Zador, A. (2022). The application of artificial intelligence to biology and neuroscience. Cell *185*, 2640-2643.

Roe, A.W., Chelazzi, L., Connor, C.E., Conway, B.R., Fujita, I., Gallant, J.L., Lu, H., and Vanduffel, W. (2012). Toward a unified theory of visual area V4. Neuron *74*, 12-29.

Rust, N.C., and Movshon, J.A. (2005). In praise of artifice. Nature Neuroscience *8*, 1647-1650.

Schrimpf, M., Kubilius, J., Hong, H., Majaj, N.J., Rajalingham, R., Issa, E.B., Kar, K., Bashivan, P., Prescott-Roy, J., and Geiger, F. (2020a). Brain-score: Which artificial neural network for object recognition is most brain-like? BioRxiv, 407007.

Schrimpf, M., Kubilius, J., Lee, M.J., Ratan Murty, N.A., Ajemian, R., and DiCarlo, J.J. (2020b). Integrative Benchmarking to Advance Neurally Mechanistic Models of Human Intelligence. Neuron *108*, 413-423.

Seidemann, E., Chen, Y., Bai, Y., Chen, S.C., Mehta, P., Kajs, B.L., Geisler, W.S., and Zemelman, B.V. (2016). Calcium imaging with genetically encoded indicators in behaving primates. Elife *5*.

Smilkov, D., Thorat, N., Kim, B., Viégas, F., and Wattenberg, M. (2017). Smoothgrad: removing noise by adding noise. arXiv preprint arXiv:170603825.

Tang, R., Song, Q., Li, Y., Zhang, R., Cai, X., and Lu, H.D. (2020). Curvature-processing domains in primate V4. Elife *9*.

Tanigawa, H., Lu, H.D., and Roe, A.W. (2010). Functional organization for color and orientation in macaque V4. Nat Neurosci *13*, 1542-1548.

Ukita, J., Yoshida, T., and Ohki, K. (2019). Characterisation of nonlinear receptive fields of visual neurons by convolutional neural network. Scientific Reports *9*.

Ustyuzhaninov, I., Burg, M.F., Cadena, S.A., Fu, J., Muhammad, T., Ponder, K., Froudarakis, E., Ding, Z., Bethge, M., and Tolias, A.S. (2022). Digital twin reveals combinatorial code of non-linear computations in the mouse primary visual cortex. bioRxiv, 2022.2002. 2010.479884.

Walker, E.Y., Sinz, F.H., Cobos, E., Muhammad, T., Froudarakis, E., Fahey, P.G., Ecker, A.S., Reimer, J., Pitkow, X., and Tolias, A.S. (2019). Inception loops discover what excites neurons most using deep predictive models. Nat Neurosci *22*, 2060-2065.

Wang, L., and Yoon, K.-J. (2021). Knowledge distillation and student-teacher learning for visual intelligence: A review and new outlooks. IEEE Transactions on Pattern Analysis and Machine Intelligence.

Willeke, K.F., Restivo, K., Franke, K., Nix, A.F., Cadena, S.A., Shinn, T., Nealley, C., Rodriguez, G., Patel, S., and Ecker, A.S. (2023). Deep learning-driven characterization of single cell tuning in primate visual area V4 unveils topological organization. bioRxiv, 2023.2005. 2012.540591.

Yamins, D.L., and DiCarlo, J.J. (2016). Using goal-driven deep learning models to understand sensory cortex. Nat Neurosci *19*, 356-365.

Yamins, D.L., Hong, H., Cadieu, C.F., Solomon, E.A., Seibert, D., and DiCarlo, J.J. (2014). Performance-optimized hierarchical models predict neural responses in higher visual cortex. Proc Natl Acad Sci U S A *111*, 8619-8624.

Zeki, S.M. (1973). Color Coding in Rhesus-Monkey Prestriate Cortex. Brain Res *53*, 422-427.

Zhuang, C., Yan, S., Nayebi, A., Schrimpf, M., Frank, M.C., DiCarlo, J.J., and Yamins, D.L.K. (2021). Unsupervised neural network models of the ventral visual stream. Proc Natl Acad Sci U S A *118*.

# Methods

### Animal preparation and surgery

All experimental protocols followed the guidelines provided by the Institutional Animal Care and Use Committee (IACUC) of Peking University Laboratory Animal Center and were approved by the Peking University Animal Care and Use Committee (LSC-TangSM-3). Three adult male rhesus macaques (Macaca mulatta) named A, B, and C, aged between 4 and 6 years, were used in this study.

The details of animal preparation for long-term calcium imaging in awake macaque have been described previously (Li et al., 2017). In summary, each animal underwent three sequential surgeries while under general anesthesia. These surgeries involved the implantation of head posts implant, virus injection, and the installation of imaging window. In the second surgery, a 20 mm craniotomy was made on the skull over the dorsal V4 region, targeting the area encompassing the lunate sulcus (lu) and the terminal portion of the inferior occipital sulcus (io). We then performed pressure injections of AAV1.hSynap.GCaMP5G.WPRE.SV40 (AV-1-PV2478, titer 2.37e13 GC/mL, Penn Vector Core) at 20-30 locations within the V4 cortex. These injections were administered at a depth of approximately 350 µm. To ensure uniform expression of GCaMP, the injection sites were spaced approximately 1 mm apart. Each injection had a volume of 100-150 nL.

### Behavioural task

Monkeys were securely restrained using a head fixation apparatus and performed an eye fixation task during image recording. The animal was required to maintain fixation on a small white spot, measuring 0.1°, within a window of 1° for 1.5 seconds to obtain a juice reward. Eye position was monitored with an infrared eye-tracking system (ISCAN) at 120 Hz.

### Visual stimuli

Visual stimuli were generated using a ViSaGe system (Cambridge Research Systems) and displayed on a 17-inch LCD monitor (Acer V173, 80 Hz refresh rate), positioned 45 cm away from the subject's eyes. Each stimulus was presented for 0.5 seconds following a pre-fixation period of 0.9 seconds. For each monkey, the receptive field of the imaged region was first estimated with small gratings. In the subsequent experiments, the stimuli were presented over the region's receptive field (Figure S2B).

**Stimuli for V4 large dataset.** The natural image stimuli used in the large dataset were sourced from ImageNet (Deng et al., 2009), specifically ILSVRC2012 and 8 synsets from the person subtree. The original images were cropped, resized and masked to create round patches measuring 180 pixels (6 degrees) in diameter with soft fade-off.

Our large-scale V4 dataset compromises a training set consisting of neural responses with each image repeated once, and a validation set consisting of neural responses with each image with ten repetitions in random interleave. Each monkey's training set contained cortical responses to over 17,000 unique color natural images. Monkey A, B, and C were tested with 20,000, 17,900, and 19,900 images, respectively, over a period of 4 to 5 days. For validation purposes, a separate set of 500 natural images was used. In addition, the validation stimuli included 48 gratings and 8 color patches (Figure S2B). The 48 gratings comprised 8 orientations (22.5° increments), 3 spatial frequencies (1.0, 2.0, and 4.0 cycles/degree), and 2 phases. The 8 color patches consisted of red, orange, yellow, green, blue, purple, white, and black. The validation sets were acquired over one day or two consecutive days. To assess the image quality and consistency of cortical responses across recording days, we generated a fingerprint stimulus set comprising the first 100 pictures from the validation stimulus set. On days when validation data was not collected, we recorded the cortical responses to these 100 fingerprint stimuli with 5 repetitions. This allowed us to evaluate the consistency of the cortical pixels' tunings as well as their imaging quality.

**Stimuli for testing the preference map.** The test stimulus set includes preferred images for multiple cortical sites. We manually selected 30 and 50 representative cortical sites with distinct feature preferences for monkey B and monkey C, respectively. For each site, we used the model's predictions to identify the top 20 images from the 50,000-image set, forming the test stimulus set. We eliminated duplicate images that were selected for different sites, resulting in stimulus sets containing 537 and 905 stimuli for monkey B and monkey C, respectively. Neural responses to these stimuli were recorded using widefield imaging and two-photon imaging techniques. During widefield imaging, the response to each stimulus was measured eight times. In the case of two-photon imaging, the response to each stimulus was measured 6-8 times within each field of view (FOV). To validate the functional domains identified by the model, we selected the model's top 16 predicted images for each domain from the test stimulus set (Figure S5A). These images were then used to test the cortical preferences and assess their alignment with the model's predictions.

**Stimuli for testing the critical feature.** We selected 12 images that were preferred by the face domain of monkey C from the above stimulus set for testing the preference map. For each of these selected images, we computed a heatmap to identify the critical region within the image that drove the response of the face domain. Using Adobe Photoshop, we created two types of images: one in which the critical region was occluded and another in which only the critical region was visible while the rest of the image was occluded. These additional images, along with the original images, formed a test stimulus set comprising a total of 36 images (12 images × 3 types). We chose a mask color to ensure a smooth transition between the occluded and unoccluded regions. The cortical responses to these stimuli were recorded using widefield imaging, with each stimulus repeated eight times.

**Widefield calcium imaging**

**Widefield imaging setup.** We performed widefield calcium fluorescence imaging with a camera imaging system adapted from Imager 3001/M (Optical Imaging). An excitation blue light was obtained with a LED light source (S3000, Nanjing Hecho Technology Co.) passing through a 470/40 nm filter. The reflected light was collected using a pair of lens (Sigma, 50mm) and split into green and blue light with a dichroic mirror (525 nm). The green light was further filtered (525/50nm) and projected onto a green channel camera (Imager 3001/M, Optical Imaging) and was recorded as the fluorescence calcium images at a rate of 33 Hz. The blue light was projected onto a blue channel camera (ZWO ASI533MC Pro, ZWO) and was recorded as the reflectance image at a rate of 20Hz (Figure 1A). The reflectance image, which captured the blood vessels well, served as the reference for anatomical registration. The imaging focus was adjusted to 300μm below the cortical surface. A fast-mechanical shutter was inserted into the blue excitation light pathway to provide intermittent illumination for long-term stable imaging.

**Calculating cortical responses.** During each stimulus presentation epoch, there were two periods of illumination, each lasting 250 ms. We utilized the fluorescence images recorded during these periods to calculate cortical responses. The first illumination period commenced at 150ms prior to the onset of the stimulus, while the second illumination period began 350ms after the stimulus onset. We averaged the frames captured during these two periods to obtain the baseline image (F0) and peak response image (F1) respectively. From these images, we computed maps of fluorescence change (ΔF/F0 map) using the formula (F1-F0)/F0. To eliminate global signal changes unrelated to the stimulus and to reveal local modulation, we performed a subtraction operation, subtracting a Gaussian blurred (σ=1.0mm) version of the ΔF/F0 map from the original map, resulting in cortical responses that highlight local changes. We found that the cortical responses obtained by this high-pass filtering matched better the responses measured by two-photon imaging as shown in Table S2.

**Image registration across days.** Prior to imaging each day, we carefully adjusted the camera system to ensure alignment with the positions on the first day. Our procedure involves focusing on the plane that exhibits the highest sharpness in the blood vessel image. We then laterally moved the camera system to match the vessel image with the reference image acquired on the first day. To assess alignment accuracy, we developed a custom MATLAB software that evaluates image sharpness and lateral position error. Once the alignment is achieved, we move the depth of focus down by 300 μm for recording.

To correct for minor displacement and distortion of the cortex across multiple imaging days, we incorporated additional image correction during data processing. We first aligned the blood vessel image captured by the blue imaging camera with the fluorescence images acquired by the green imaging camera. This alignment involved rotating, rescaling, and translating the blood vessel images. Then, a transformation matrix was generated between the blood vessel images from each day and the reference image acquired on the first day. This transformation matrix was used to correct and align the fluorescence images across days. Since the blood vessel images acquired by the blue channel camera have higher spatial resolution than the fluorescence images, this approach allows us to achieve greater registration accuracy.

**DNN Modeling**

**Data preprocessing for modeling.** The stimulus images, initially measuring 200×200 pixels (30 pixels/degree), were resized to 100×100 pixels and input to the model. The raw response map acquired by the camera had a resolution of 512×512 pixels and a sampling rates of 45 pixels/mm, which exceeded the intrinsic spatial resolution of the widefield calcium imaging signal. To simplify the modeling analysis, we rescaled the response maps to 128×128 pixels using bilinear interpolation. We used validation set responses to identify regions exhibiting significant stimulus-related fluorescence changes. Specifically, we conducted one-way ANOVA across responses of 556 validation stimuli, resulting in regions with p-values below $10^{-300}$, indicating highly significant changes. These regions were designated as regions of interest (ROIs, Figure S2A) for modeling purpose. Responses of regions from outside the ROIs were masked to zero during modeling.

**Network architecture.** Our feature transfer model consisted of a feature extraction and a two-layer perceptron (Figure S3A). The features are the output maps of Add-6 layer of Xception (Chollet, 2017) to a stimulus image. We used the Add-6 layer as its outputs have been demonstrated to be a reliable predictor of V4 responses (see Figure S3B). We used a Keras implementation of Xception, which was trained for the ImageNet classification task. The two-layer perceptron was used to map the features to the corresponding cortical responses. The hidden layer of the perceptron consists of 200 units, designed to extract effective features while avoiding overfitting. To enhance the model's expressive capacity, we incorporate an exponential linear unit (ELU) nonlinearity in the hidden layer. This nonlinearity aids in capturing complex relationships and enhancing the model's ability to represent the data.

The small Xception-like DNN model (Figure S3C) consisted of a CNN encoder that shared architectural similarities with Xception, including depth-wise separable convolution layers and residual learning blocks. The encoder generated nonlinear feature responses from input images. A readout network was used to map the output of encoder to cortical responses. Sigmoid activation function was used to introduce nonlinearity. The encoder converted the RGB input images to 7×7×400 feature maps, which were fed into the readout network. Each position within the feature maps corresponded to a distinct retinotopic spatial location, containing a column of features, whereas different positions in the imaged V4 cortex encoded different features with similar spatial receptive fields. In order to transform the spatially organized maps of the encoder into feature-organized maps that resembled the organization in the cortex, the readout network first reorganized the input 7×7×400 spatial-feature map into a 20×20×49 feature-spatial map. This reorganization swapped the spatial and feature organization. The 400 feature channels were now organized as a 20×20 spatial map, with each column containing that feature channel's responses across various retinotopic locations. This feature-spatial map was then passed through sequences of convolutional and locally connected layers to generate the final response output.

To prevent overfitting, dropout layers with a dropout ratio of 0.1 were introduced to all of the above models, meaning that during training, 10% of the responses were randomly dropped to encourage the model to generalize better to unseen data. Additionally, the feature transfer model utilized an L1 penalty regularization on the connection weights between the hidden layer and Xception features. This regularization encourages sparsity and promotes the selection of more relevant and informative features, reducing the risk of overfitting and improving generalization.

**Model training.** All the models were optimized using stochastic gradient descent with the Adam optimizer (Kingma and Ba, 2014) and a batch size of 20. When fitting the neural data, models are trained to minimize the mean square error (MSE) on the training set. To prevent overfitting and ensure the best generalization performance, we used early stopping based on the MSE between predicted and measured neural responses on the validation set. If the MSE failed to decrease during any consecutive 50 passes through the entire training set (50 epochs), the training process would be halted. The model that achieved the best performance on the validation set during the training phase would be saved as the final model. This approach allows us to capture the model's optimal performance while avoiding unnecessary training iteration.

For knowledge distillation, we used the results generated by the feature transfer model as data to train the small DNN model using supervised training. This generated data consisted of image-response pairs derived from 100,000 ImageNet images. The responses were predicted by the feature transfer model that was trained on the neural data. The generated data was split into a training set containing 90,000 images and a validation set containing 10,000 images. We trained the small DNN on this data by minimizing the training set MSE. We also used early stopping based on the MSE on the validation set. The training would be halted if the MSE on the validation set did not decrease in ten consecutive epochs.

**Predictivity evaluation.** The models were assessed by evaluating the correlation between the predicted responses and the measured responses of each cortical pixel to the validation stimuli. There is an inherent limit to the maximum achievable correlation due to response variability within the same day and across days. To estimate this limit, or noise ceiling, we calculated the correlation (see Figure S2F for statistics) between the responses to the fingerprint stimuli in the validation set and the responses to the fingerprint stimuli averaged across days when the training set were collected. This allowed us to determine the upper bound of correlation that can be achieved considering the inherent variability in the neural responses. The models' achieved correlation is then normalized by the noise ceiling to provide a performance measure.

## Preference map analysis

**Preference map synthesis.** We gathered a set of 50,000 images from ImageNet and prepared them in the same way as we did for generating the stimulus sets used in the experiment. We used the KD Transfer model to predict the cortical responses of these 50,000 images. We then organized the nine most responsive images for each cortical pixel into a 3×3 grid and display it at the corresponding cortical location in the acquired image to derive the preference map.

**Hierarchical clustering on the preference map.** In the preference map, the preference of each cortical pixel is represented by its top nine images. To cluster these cortical pixels based on the similarities of their preferred images, we employed a method illustrated in Figure S4. First, we computed the model's prediction of each cortical pixel's response to the entire set of 50,000 images, and normalized them to range between 0 and 1. Then, we combined all the pixels' normalized responses to generate a predicted cortical activation pattern associated with each image. We then averaged the activation patterns of the top nine images preferred by each pixel to create the 'cortical response vector' of that pixel. The similarity between two cortical pixels was determined by computing the Pearson correlation between their respective cortical response vectors. This similarity metric enabled us to identify groups of cortical pixels that exhibited similar activation patterns when exposed to their preferred images. To identify functional domains within the V4 cortex based on shared image preferences, we employed hierarchical clustering based on average-linkage, and grouped cortical pixels within a distance threshold of 0.4 (computed as one minus the similarity) into a cluster. Any cluster that included connected regions larger than 40 pixels was considered a functional domain. This approach allowed us to distinguish distinct functional domains within the V4 cortex, based on their collective preference for specific image features.

## Two-photon calcium imaging

**Two-photon imaging setup.** We performed two-photon calcium imaging on monkeys B and C with a Bruker two-photon imaging system (Prairie Ultima IV, Bruker Nano). The wavelength of the femtosecond laser (Insight X3, Spetra-Physics) was set to 1000 nm. Field of views (FOVs) of 600 μm × 600 μm were imaged under 1.4× zoom with a 16× objective (0.8-N.A., Nikon) at a resolution of 1.2 μm/pixel. A fast-resonant scan (30 frames per second) was used to obtain images of neuronal activity. We averaged every two frames, resulting in an effective frame rate of 15 fps. In total, we recorded 20 FOVs, 10 from monkey B and 10 from monkey C, with recording depths ranging from 100 μm to 300 μm. To determine the precise position of each FOV relative to the widefield imaging map, we recorded the blood vessel image directly above each FOV as a reference to align the two-photon imaging data with the widefield imaging map.

**Data processing for two-photon imaging.** We used customized MATLAB code to process the data obtained from the experiments. First, we associated the two-photon image series with the corresponding visual stimuli using the time sequence information recorded by Neural Signal Processor (Cerebus system, Blackrock Microsystem). Then, the images were motion corrected using a normalized cross-correlation-based translation algorithm (Li et al., 2017). This step helped to align the images and mitigate any image shifts caused by motion during recording. For the response to each stimulus, we computed the F0 image by averaging the five frames preceding the onset of the stimulus. Similarly, the F1 image was obtained by averaging the frames from the fifth to the tenth frames after stimulus onset. These F0 and F1 images provided baseline and peak response information associated with a stimulus, respectively. An additional non-rigid motion correction (Giovannucci et al., 2019) was applied to the F0 and F1 images to correct for the cortical deformation during the long recording session.

We used the differential image (ΔF) obtained by F1-F0 to extract regions of interest (ROIs). We averaged the differential images across all repeated trials of the same stimulus. A band-pass Difference of Gaussian filtering (standard deviations of positive and negative Gaussians are 1 and 30 pixels respectively) was then applied to the averaged differential images. The connected subsets of pixels (>30 pixels) with pixel values> 3.5 standard deviations of the mean brightness were selected as ROIs. We further refined the shape of the ROI by calculating the correlation between the ΔF values of the ROI and its neighboring pixels. Pixels with a correlation greater than 0.3 will be assigned to the ROI. Using the above methods, we obtained many overlapping ROIs. To determine whether these ROIs should be merged, we perform hierarchical clustering on the responses of pixels within these ROIs. The pixel response was calculated by ΔF/F of the ROI to which it belonged. In cases where a pixel belonged to the multiple ROIs, the response was computed as the average of ΔF/F of those ROIs. We calculated the response of each selected ROI using ΔF/F and identified visually responsive ROIs by discarding ROIs that did not exhibit significant response selectivity

($p>10^{-5}$, tested with a one-way ANOVA) for any of the 537 and 905 test stimuli for monkey B and monkey C, respectively.

**Heatmap analysis**

**Heatmap synthesis.** To produce a heatmap for a specific image of interest and a target cortical region, we employed the SmoothGrad-Square algorithm (Hooker et al., 2019; Smilkov et al., 2017). This algorithm relies on computing the gradient map of the model's response with respect to the input image. The SmoothGrad-Square algorithm operates by introducing Gaussian noise to the image of interest, generating a set of similar images. For each generated image, the model's response is backpropagated to the input of the deep learning model to generate a gradient map, which captures the sensitivity of the model's response to changes in the input. These gradient maps are then squared and aggregated to produce the final heatmap. SmoothGrad-Square involves two hyper-parameters: σ, the standard deviation of the Gaussian noise, and n, the number of samples to sum over. Here we used σ = 0.2 (image value $\in [0,1]$) and n = 20.

**Estimating receptive field.** To estimate the receptive field of the target cortical region, we averaged the heatmaps generated from a large set of natural images. Namely, we first calculated the heatmaps for the top 1,000 images in the 50,000-image set for the target cortical region. Next, we normalized the heatmap for each image to ensure that the sum of values on the heatmap equals the cortical response elicited by the image. Finally, we fitted the average of these normalized heatmaps with an elliptical Gaussian:

$$f(x, y) = A \cdot \exp\left(-\frac{x'^2}{2\sigma_a^2} - \frac{y'^2}{2\sigma_b^2}\right) + B$$

$$\begin{cases} x' = (x - x_c)\cos\theta + (y - y_c)\sin\theta \\ y' = -(x - x_c)\sin\theta + (y - y_c)\cos\theta \end{cases}$$

where $A$ is the amplitude of the Gaussian, $B$ is the offset, $\sigma_a$ and $\sigma_b$ are the standard deviations of the elliptical Gaussian along its two principal axes, and $x'$ and $y'$ are transformations of the coordinates $x$ and $y$, taking into account the angle $\theta$ and the offset ($x_c$ and $y_c$) of the ellipse. In total, there were seven free parameters in the fitting procedure: $A, B, \sigma_a, \sigma_b, \theta, x_c$ and $y_c$. We define the square root of the area within 1 SD (i.e. $\sqrt{\pi\sigma_a\sigma_b}$) as RF size.

# Supplementary Information

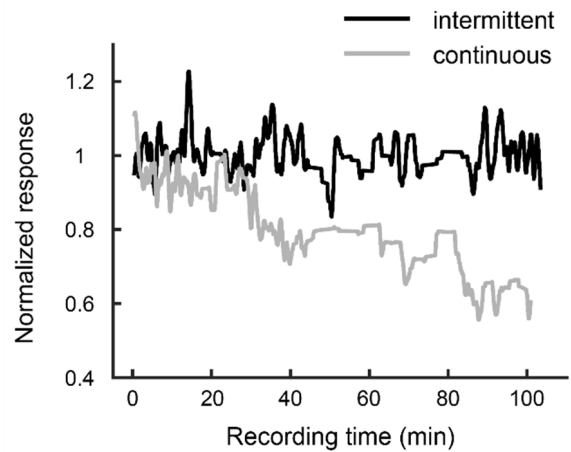

**Figure S1. Comparison of photobleaching effect between continuous and intermittent illumination, Related to Figure 1**

Cortical responses to 2,000 distinct natural images were recorded under continuous illumination (grey line) and intermittent illumination (black line). The average response signal within the ROI exhibited a gradual attenuation under continuous illumination while the response remained stable under intermittent illumination. Both response curves were smoothed with a Gaussian filter (sigma = 10 trials).

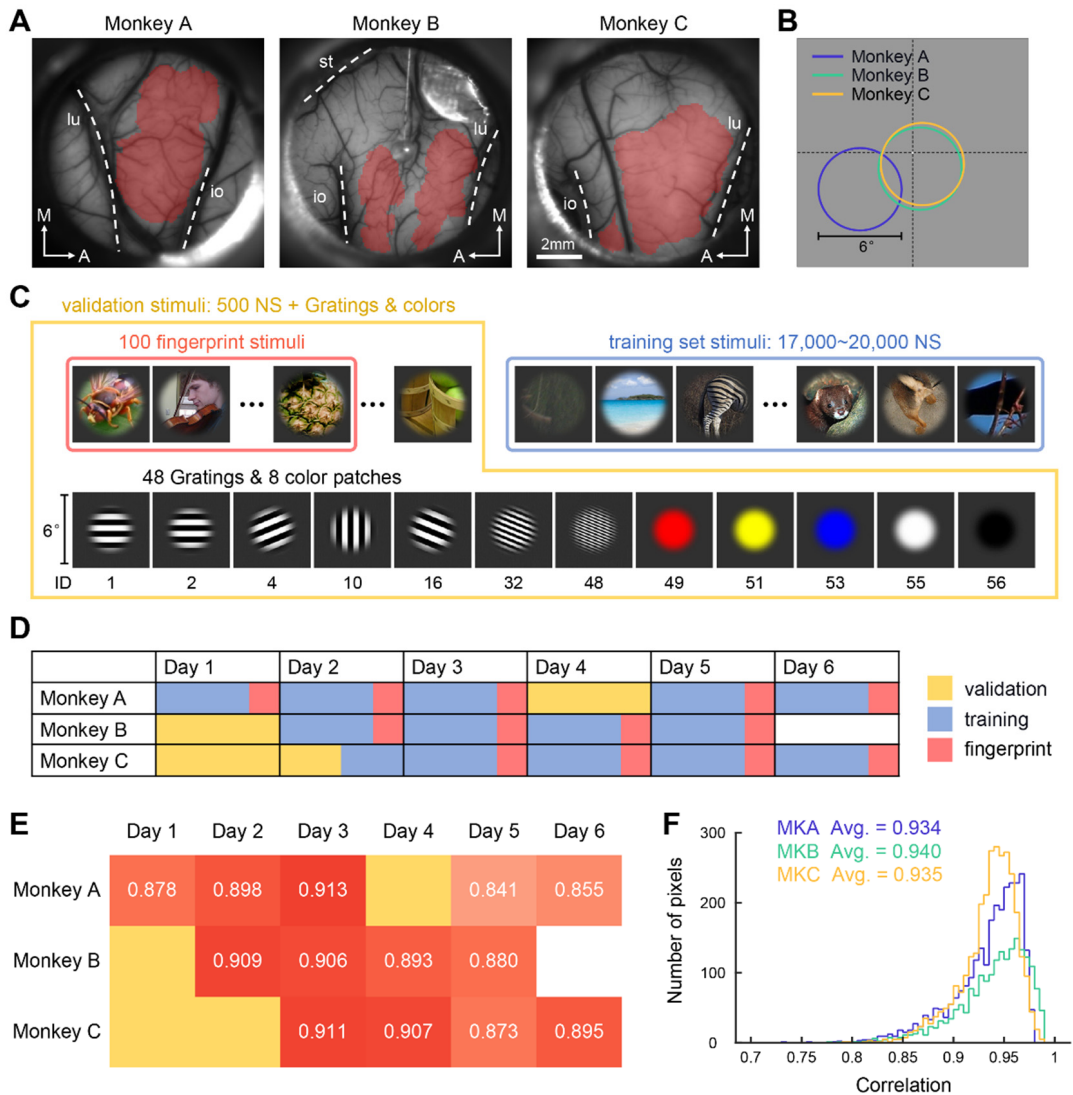

**Figure S2. Overview of the widefield calcium imaging dataset, Related to Figures 1 and 2**

(A) Optical window for each monkey. Regions of interest (ROIs) selected for modeling analysis are highlighted in red. These regions were identified by selecting cortical pixels with significant selective responses to the validation stimuli (determined through one-way ANOVA across responses of 556 validation stimuli, $p<10^{-300}$). Anatomical landmarks are labeled as follows: A, anterior; M, medial; io, inferior occipital sulcus; lu, lunate sulcus; st, superior temporal sulcus.

(B) Stimulus presentation regions for three monkeys. Before data acquisition, we used small gratings presented at various locations to map the receptive field (RF) of the imaged cortical region. A specific stimulus presentation region was centered on the mapped RF for each monkey. The fixation point is indicated by the central white dot. For monkey B and monkey C, the RFs of the imaged regions were situated near the fovea, while for monkey A, the RF was more peripheral.

(C) The stimulus set. The validation stimuli included 500 natural stimuli (NS), 48 gratings, and eight color patches. The gratings encompassed eight orientations (22.5° increments), three spatial frequencies (1.0, 2.0, and 4.0 cycles/degree), and 2 phases. The cortical responses to each validation stimulus were measured ten times. A separate set of 17,000-20,000 natural images constituted the training set. The response to a stimulus in each training set was measured once. The first one hundred natural images from the validation stimulus set were also used as fingerprint images. These images were presented with five repetitions on days when the validation data were not collected.

(D) The collection schedule of the calcium imaging dataset. The dataset for each monkey was collected over a span of 5-6 days. The validation set was recorded on the days that are marked yellow. The remaining days involved daily testing of the fingerprint stimuli (red) to monitor the stability of cortical responses.

(E) The correlation of cortical responses to fingerprint stimuli across different days. For each cortical pixel within a ROI, the Pearson correlation was computed between the trial-averaged responses of the fingerprint stimuli from the fingerprint session and the validation session. The average correlation for all ROI pixels is shown for each fingerprint session. The high level of correlations (>0.8) indicated that the cortical responses were stable during the measurement period.

(F) The distribution of the noise ceilings for cortical pixels, evaluated by calculating the correlation between the averaged responses of fingerprint stimuli across all fingerprint sessions and those in the validation session. Different colors represent different monkeys.

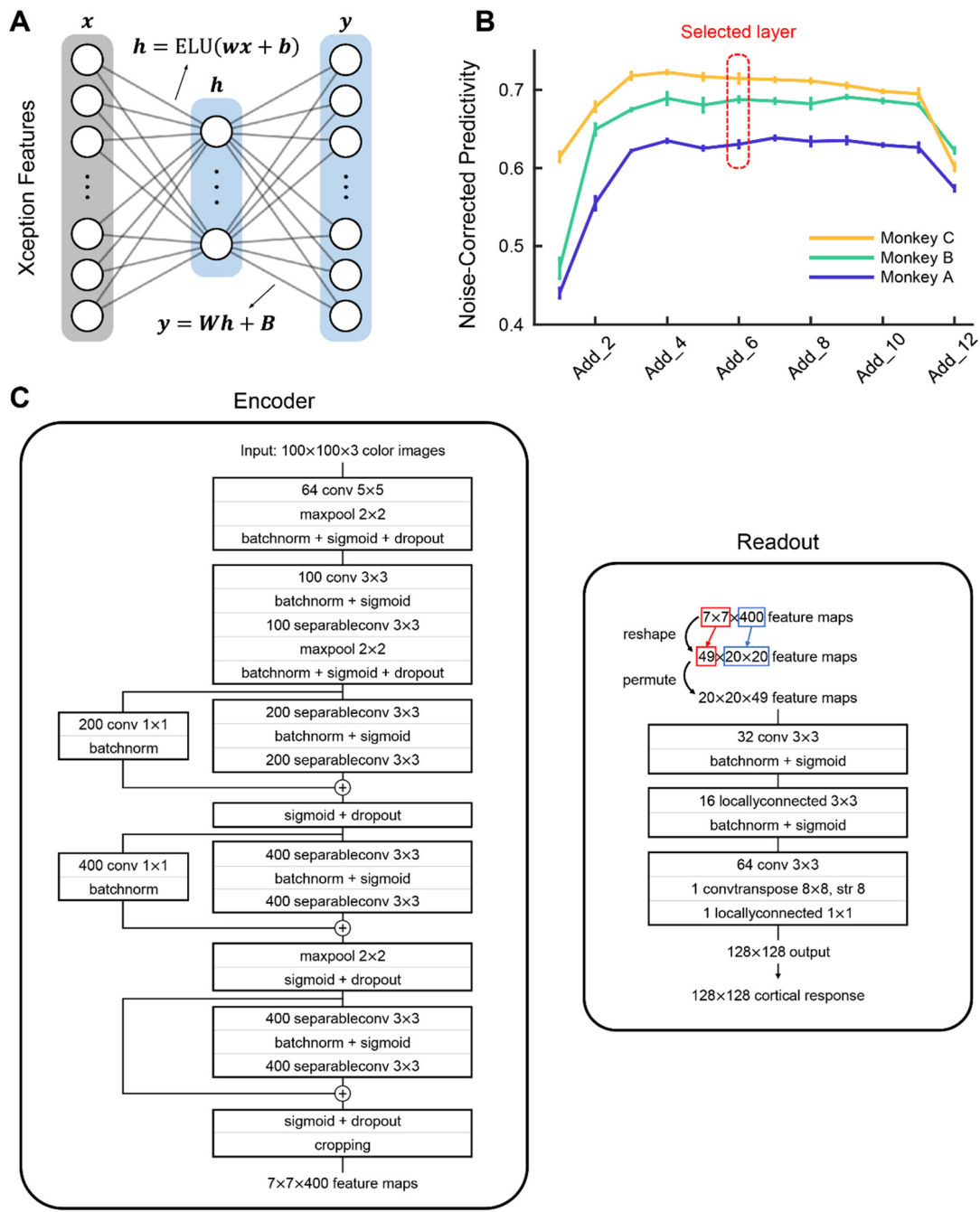

**Figure S3. Network design of the neural models, Related to Figure 2**

(A) The network architecture of the feature transfer model utilized a two-layer perceptron to map a Xception layer feature responses ($x$) to the cortical responses ($y$). To enhance the expressive power of the model, we applied an exponential linear unit (ELU) nonlinearity to the hidden layer.

(B) Selection of the Xception layer for the feature transfer model. We tested models based on the output layer of each residual block of the Xception, which corresponds to the Add layer. The Add_6 layer (marked by the red box) was selected as it provided a feature map for our feature transfer model that exhibited good performance across all monkeys' data. The error bar represents the standard deviation across different initializations.

(C) The network architecture of the small Xception-like DNN consisted of an encoder and a

readout network. The encoder was responsible for generating nonlinear features map from input images, while the readout network mapped the feature responses to cortical responses. The encoder shared architectural similarities with Xception, including sequences of convolution layers (indicated by 'conv' with values denoting feature depth and convolution filter size), batch-normalization layers ('batchnorm'), depth-wise separable convolution layers ('separableconv'), max-pooling layers ('maxpool') and residual learning blocks. The readout network transformed the spatially organized maps of the encoder into features-organized cortical response maps. This was achieved by first reorganizing the 7×7×400 spatial-feature map into a 20×20×49 feature-spatial map. The map was then passed through sequences of convolution and locally connected layers ('locallyconnected'). Finally, a transposed convolution layer ('convtranspose' with 'str' denoting filter stride) transformed the map into a 128×128 response output.

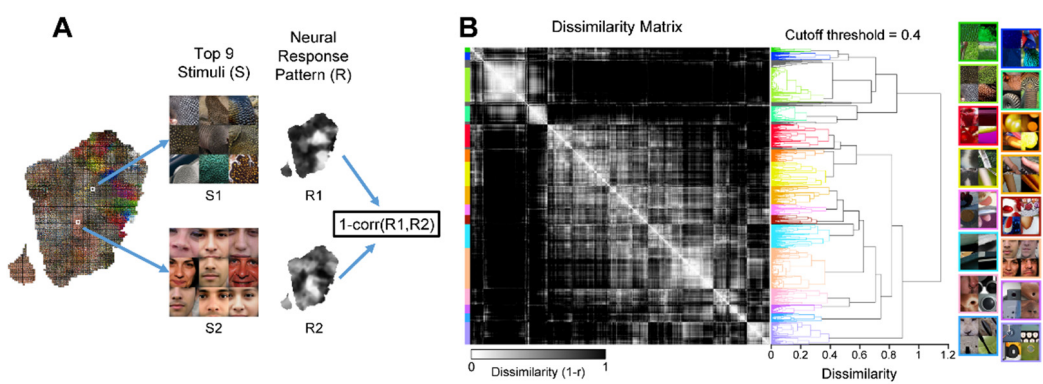

**Figure S4. Illustration for hierarchical clustering on preference map, Related to Figure 3**

(A) The similarity distance between cortical pixels is computed as follows. Each cortical pixel's preference is represented by its top nine images. The model-predicted response patterns of these images are averaged, and the dissimilarity between a pair of cortical pixels is computed based on the correlation distance between their response patterns.

(B) Using the dissimilarity measure above, we performed hierarchical clustering to identify the functional domains. Left: The matrix indicates the dissimilarity between pairs of cortical pixels. Pixels were sorted based on their order in the hierarchical clustering. Middle: The dendrogram of the hierarchical clustering process. A cutoff threshold of 0.4 was used to identify the functional domains. Right: The images predicted to evoke strong responses for each identified domain. The functional domains are represented by different colors in the color scheme, which indicates their respective categories.

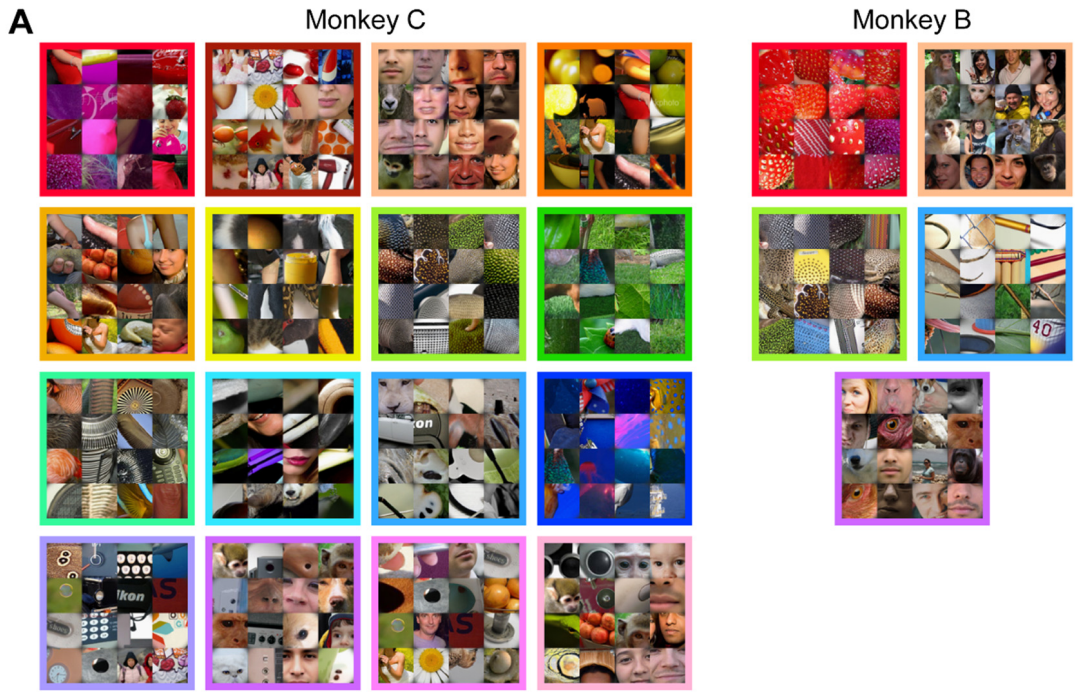

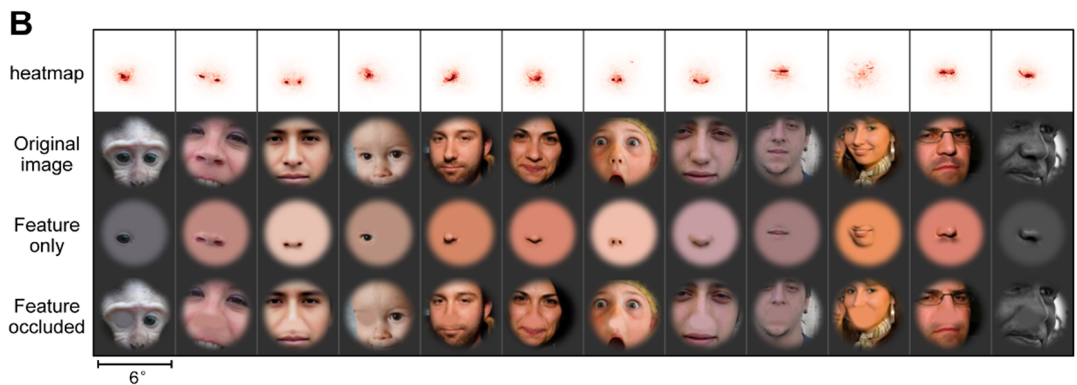

**Figure S5. Stimuli for *in vivo* testing of the model prediction, Related to Figures 4 and 6**

(A) Stimuli for testing the model-predicted functional domains. We selected 16 stimuli for each domain that were predicted to evoke strong responses. For the purpose of visualization, only the central 4×4 degrees of each stimulus are shown.

(B) Stimuli for testing the face domain's critical feature in monkey C. The first row displays the heatmaps for the original natural images (second row), the third row contains images with only the critical features visible, and the last row comprises images with critical features occluded.

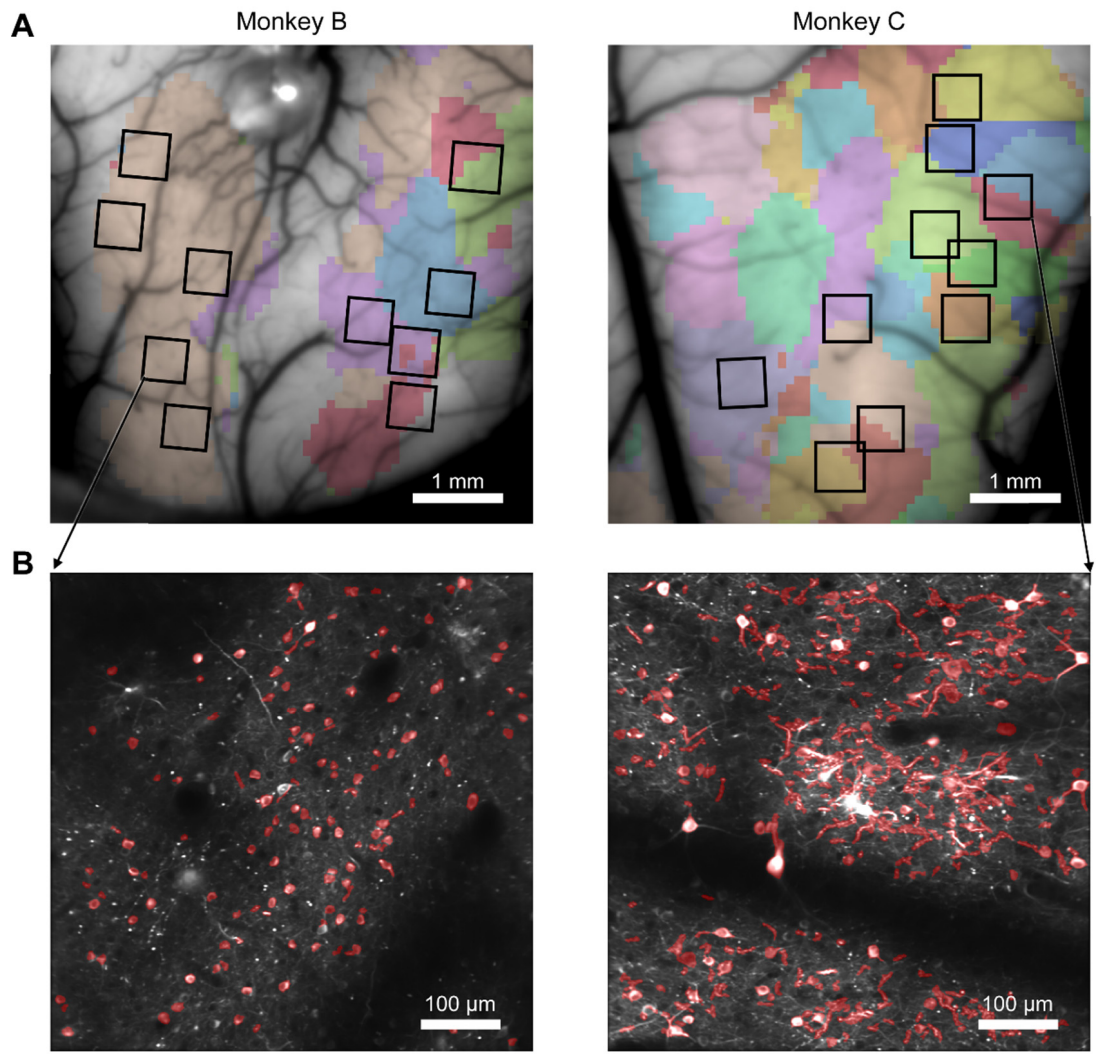

**Figure S6. Two-photon imaging fields of view (FOVs), Related to Figure 5**

(A) The recording sites in monkey B and monkey C. The color hue represents the measured cortical preference, following the same color scheme as in Figure 4D.

(B) Two example FOVs from monkey B and monkey C. The identified visually responsive regions of interest (ROIs) were marked in red, which included both soma and dendrites.

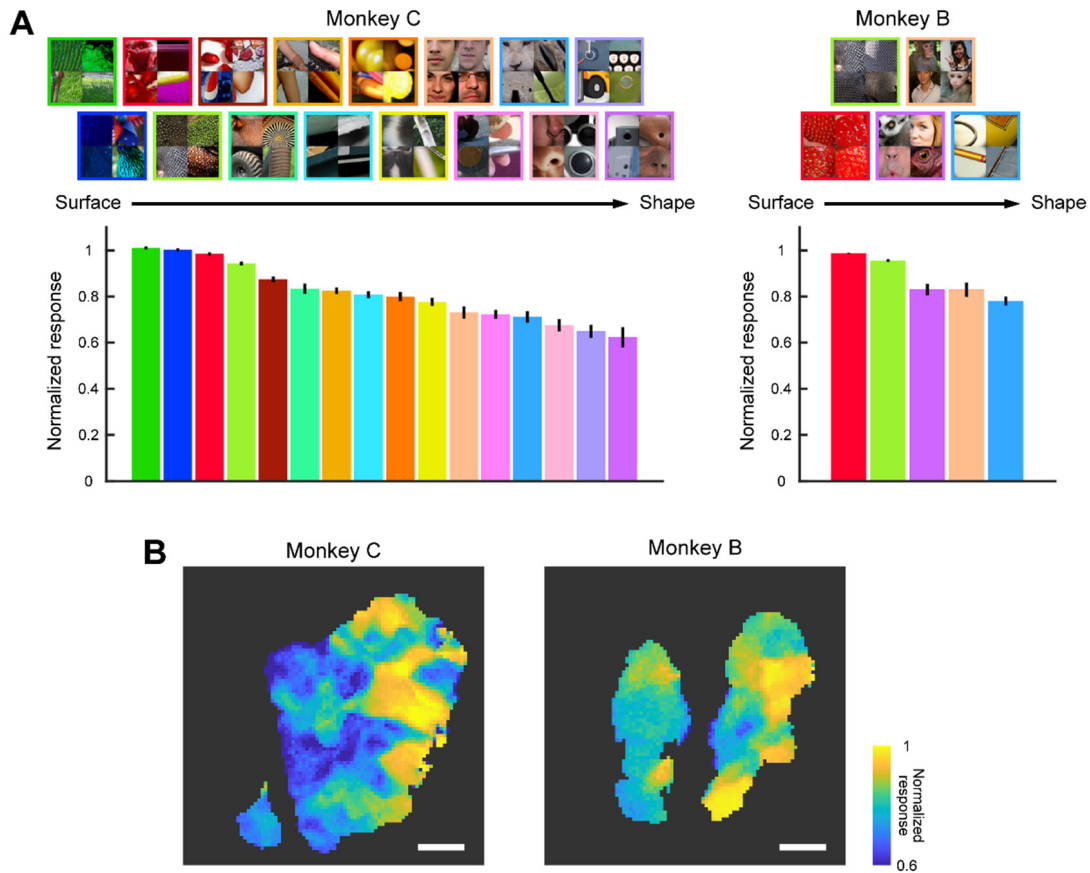

**Figure S7. Assessing the cortical feature preference between shape and surface attributes, Related to Figure 6**

(A) Domains preferring shape attributes exhibit a more significant response decrease under small perturbation (rate = 10%). The top shows the preferred images for each domain (monkey C data same as Figures 6G). The bottom shows the domains' normalized responses to perturbed images. The value and error bar show the average and SEM of each domain's responses to the perturbed images, which are derived from the domain's top 25 images.

(B) Maps for surface-shape preference. For each cortical pixel, preferences between surface and shape features were measured by cortical responses under small perturbation (rate = 10%, as in (A)). The clusters in the maps suggest a potential separate organization of surface and shape features in V4. The scale bar denotes 1 mm.

| Model | Monkey A | Monkey B | Monkey C | Total |
|---|---|---|---|---|
| Fine-tuning | 0.638±0.002 | 0.674±0.002 | 0.703±0.002 | 0.672±0.001 |
| Feature Transfer | 0.630±0.002 | 0.695±0.002 | 0.722±0.002 | 0.682±0.001 |
| Direct | 0.635±0.002 | 0.691±0.002 | 0.733±0.002 | 0.688±0.001 |
| KD Transfer | **0.679±0.002** | **0.745±0.002** | **0.769±0.001** | **0.731±0.001** |

**Table S1. Noise-corrected predictivity of the cortical response models, Related to Figure 2**
The Fine-tuning model and Feature Transfer model share the same network architecture, consisting of feature response computation and a two-layer perceptron readout. The difference between them is that the network weights for generating the feature responses in the Feature Transfer model are fixed, only the parameters of the two-layer perceptron are trained using neural data. In contrast, both the feature extraction and two-layer perceptron are trainable in the Fine-tuning model. The Fine-tuning model initialized the weights of its feature extraction network using the weights from pre-trained Xception. Similarly, the KD Transfer model and the Direct training model share the same architecture. The discrepancy arises in the weight initialization when fitting the neural data. The Direct training model initializes its weights randomly, whereas the KD Transfer model obtains its initial weights through knowledge distillation from the Feature Transfer Model. The reported values and errors in the table represent the average and standard error of the mean (SEM) across all cortical pixels from each monkey. Bold indicates the best model performances.

| Filter parameter | Monkey B | Monkey C |
|---|---|---|
| σ = 0.2 mm | 0.650±0.036 | 0.655±0.033 |
| σ = 0.5 mm | 0.709±0.032 | 0.722±0.027 |
| σ = 1.0 mm | **0.748±0.030** | **0.759±0.019** |
| σ = 2.0 mm | 0.739±0.032 | 0.752±0.019 |
| σ = 5.0 mm | 0.739±0.028 | 0.675±0.025 |
| none | 0.727±0.029 | 0.609±0.028 |

**Table S2. Various filtering parameters were tested for calculating cortical responses in widefield imaging, Related to Methods**
We computed the correlation between the responses of the neuronal population in a two-photon imaging field of view and the responses of the corresponding area in widefield imaging. Our results showed that the optimal filter parameters that yielded the best match between the cortical responses and the neuronal population responses are σ = 1.0 mm. The reported values and error represent the average correlation and standard error of the mean (SEM) across the two-photon imaging FOVs from each monkey. Bold indicates the best match.